\pdfoutput=1
\documentclass[ALICE,manyauthors]{cernphprep}

\usepackage[comma,square,numbers,sort&compress]{natbib}
\usepackage{bm}
\usepackage{hyperref}
\usepackage{lineno}
\graphicspath{{Results/Figures/}}
\usepackage{color}
\usepackage[normalem]{ulem}
\usepackage[inline]{enumitem}
\usepackage{siunitx}
\usepackage[T1]{fontenc}
\newcommand {\pp}        {\ensuremath{\mathrm {pp}}}
\newcommand {\PbPb}      {\ensuremath{\mathrm{Pb\mbox{--}Pb}}}

\newcommand {\jpsi}      {\ensuremath{\mathrm{J}\kern-0.02em/\kern-0.05em\psi}}
\newcommand {\psip}      {\ensuremath{\psi\mathrm{(2S)}}}

\newcommand {\pt}        {\ensuremath{p_{\mathrm{\textsc{t}}}}}

\newcommand {\meanpt}    {\ensuremath{\langle p_{\mathrm{\textsc{t}}} \rangle}}
\newcommand {\meanptsq}  {\ensuremath{\langle p_{\mathrm{\textsc{t}}}^{2} \rangle}}

\newcommand {\y}         {\ensuremath{y}}

\newcommand {\sqrtsnn}           {\ensuremath{\sqrt{s_{_{\mathrm{NN}}}}}}

\newcommand {\sqrtsnnE}[2][TeV]  {$\sqrtsnn = #2\;\mathrm{#1}$}

\newcommand {\sqrts}             {\ensuremath{\sqrt{s}}\,}
\newcommand {\sqrtsE}[2][TeV]    {$\sqrts = #2\;\mathrm{#1}$}
\newcommand{\snn}{\ensuremath{\sqrt{s_{\mathrm{NN}}}}\xspace}
\newcommand {\Raa}       {\ensuremath{R_\mathrm{AA}}}
\newcommand {\rAA}       {\ensuremath{r_\mathrm{AA}}}
\newcommand {\Taa}       {\ensuremath{\langle T_\mathrm{AA} \rangle}}

\newcommand {\Npart}     {\ensuremath{\langle N_{\mathrm{part}} \rangle}}

\newcommand {\gev}       {\ensuremath{\,\mathrm{GeV}}}
\newcommand {\tev}       {\ensuremath{\,\mathrm{TeV}}}
\newcommand {\mevc}      {\ensuremath{\,\mathrm{MeV}\kern-0.05em/\kern-0.02em c}}
\newcommand {\mevcc}     {\ensuremath{\,\mathrm{MeV}\kern-0.05em/\kern-0.02em c^2}}
\newcommand {\gevc}      {\ensuremath{\,\mathrm{GeV}\kern-0.05em/\kern-0.02em c}}
\newcommand {\gevcc}     {\ensuremath{\,\mathrm{GeV}\kern-0.05em/\kern-0.02em c^2}}
\newcommand {\gevcsq}    {\ensuremath{\,\mathrm{GeV}^2\kern-0.05em/\kern-0.02em c^2}}
\newcommand {\gevfmcube} {\ensuremath{\,\mathrm{GeV}\kern-0.05em/\kern-0.02em \mathrm{fm}^3}}
\newcommand {\Ae}        {\ensuremath{A \times \varepsilon}}

\begin{document}%

%%%%%%%%%%%%%%%  Title page %%%%%%%%%%%%%%%%%%%%%%%%
\begin{titlepage}
% CERN-EP-2019-187
\PHyear{2019}
\PHnumber{187}      % required, will be obtained from PH
\PHdate{2 September}  % required, will be obtained from PH
%

%%% Put your own title + short title here:
\title{Studies of \jpsi\ production at forward rapidity in Pb--Pb collisions at \mbox{\texorpdfstring{\snn}{sqrt(s\_NN)}=5.02~TeV}}
\ShortTitle{\jpsi\ production in Pb--Pb collisions at \sqrtsnnE{5.02}}   % appears on right page headers

%%% Do not change the next lines
\Collaboration{ALICE Collaboration\thanks{See Appendix~\ref{app:collab} for the list of collaboration members}}
\ShortAuthor{ALICE Collaboration} % appears on left page headers, do not change

\begin{abstract}
%%No setails on pp
The inclusive \jpsi\ production in \PbPb\ collisions at the center-of-mass energy per nucleon pair \sqrtsnnE{5.02}, measured with the ALICE detector at the CERN LHC, is reported. The \jpsi\ meson is reconstructed via the dimuon decay channel at forward rapidity ($2.5<\y<4$) down to zero transverse momentum. The suppression of the \jpsi\ yield in \PbPb\ collisions with respect to binary-scaled \pp\ collisions is quantified by the nuclear modification factor (\Raa). The \Raa\ at \sqrtsnnE{5.02} is presented and compared with previous measurements at \sqrtsnnE{2.76} as a function of the centrality of the collision, and of the \jpsi\ transverse momentum and rapidity. The inclusive \jpsi\ \Raa\ shows a suppression increasing toward higher transverse momentum, with a steeper dependence for central collisions. The modification of the \jpsi\ average transverse momentum and average squared transverse momentum is also studied. Comparisons with the results of models based on a transport equation and on statistical hadronization are carried out.

\end{abstract}
\end{titlepage}
\setcounter{page}{2}

\section{Introduction}
\label{sec:introduction}

The study of ultra-relativistic heavy-ion collisions aims to investigate the properties of strongly-interacting matter at high temperature and energy density. Lattice Quantum Chromodynamics calculations predict that a deconfined state of partonic matter, the so-called Quark--Gluon Plasma (QGP), can be created in such collisions~\cite{Bazavov:2014pvz,Borsanyi:2013bia,Braun-Munzinger:2015hba}. Among the many possible probes to study this phase of matter, heavy quarks (charm (c) and beauty (b)) are particularly interesting as they are expected to be produced in the initial stage of the collisions, by hard partonic scatterings, and to experience the full evolution of the system. In particular, it was predicted that bound states of c and $\overline{\rm c}$ quarks (known as charmonia) should be suppressed due to the color-screening mechanism~\cite{Matsui:1986dk}. The suppression probabilities of the quarkonium (c$\overline{\rm c}$ or b$\overline{\rm b}$) states in the QGP depend on their binding energies and the medium temperature. Therefore, the measurement of the relative production rates of the quarkonium states should give indications on the temperature of the system~\cite{Digal:2001ue}.  Among the different charmonium states, the study of the ground state with quantum numbers $J^{PC} = 1^{- -}$ (\jpsi) is comparatively more accessible due to its larger abundance and to the relatively large branching ratio to dileptons, and has led to several important results. 
 %The measurement of the relative production probabilities of charmonia in nuclear collisions is expected to depend on their binding energy, and gives indications on the initial temperature of the system~\cite{Matsui:1986dk,Digal:2001ue}. Among the different charmonium states, the study of the ground state (\jpsi) is comparatively easier, and has led to several important results. 

%Its production relies on the combination of prompt and non-prompt sources. 
%The prompt \jpsi\ yield is the sum of direct ($\simeq 65\%$) and excited ($\simeq 24\%$) charmonium states such as \chic\ and \psip\ decaying through the \jpsi\ + X channel~\cite{LHCb:2012af,Aaij:2012ag}. The non-prompt production is related to the decay of beauty hadrons ($\simeq 10\%$) whose relative contribution increases with the colliding energy and the transverse momentum (\pt) of the \jpsi~\cite{Aaij:2011jh}.

Over the past decades, the \jpsi\ production in heavy-ion collisions was measured at the SPS, RHIC and the LHC, covering a wide range of center-of-mass energies per nucleon pair (\sqrtsnn) from about 17 \gev\ to 5.02 \tev. A suppression of the \jpsi\ production yield in nucleus--nucleus (AA) relative to that expected from measurements in proton--proton (\pp) collisions was observed at the SPS at \sqrtsnn\ $=$ 17 \gev~\cite{Alessandro:2004ap,Arnaldi:2007zz}, at RHIC up to \sqrtsnnE{0.2}~\cite{Adare:2006ns,Adare:2011yf,Adamczyk:2016srz,Abelev:2009qaa} and at the LHC at \sqrtsnnE{2.76}~\cite{Abelev:2013ila,Abelev:2012rv,Adam:2015isa,Adam:2015rba,Chatrchyan:2012np} and 5.02 TeV~\cite{Adam:2016rdg,Sirunyan:2017isk,Aaboud:2018quy}. The suppression is evaluated through the calculation of the nuclear modification factor (\Raa), corresponding to the ratio of the production yields in AA and the cross section in \pp\ collisions, normalised by the nuclear overlap function (\Taa)~\cite{Miller:2007ri}. The observed suppression does not increase with increasing collision energy as expected in the color-screening picture considering the increasing temperature of the formed QGP. This observation is naturally explained by a further production mechanism known as regeneration, in which abundantly produced c$\overline{\rm c}$ pairs recombine into \jpsi~\cite{Thews:2000rj,BraunMunzinger:2000px}. The contribution of the regeneration to \jpsi\ production has to increase with the density of c$\overline{\rm c}$ pairs and consequently with the collision energy. It is worth noting that the regeneration contribution should favour low transverse momentum (\pt) \jpsi, as the bulk of charm quarks are produced at small momenta~\cite{Thews:2000rj,BraunMunzinger:2000px}. The regeneration scenario was further supported by the measurement of a positive \jpsi\ elliptic flow ($v_{2}$)~\cite{ALICE:2013xna,Khachatryan:2016ypw,Acharya:2017tgv,Acharya:2018pjd,Aaboud:2018ttm} which, at low \pt, can be acquired via charm-quark recombination~\cite{Du:2015wha,Liu:2009nb}. 
It is important to note that in addition to the effects discussed above, related to the production of a high energy-density medium, the so-called cold-nuclear-matter effects may also have a sizeable influence on the charmonium yields. In particular, the modification of the parton distribution functions in the nucleus (e.g. nuclear shadowing~\cite{Eskola:2016oht,Kovarik:2015cma}) may modify the initial yields of charm quarks and has to be taken into account in the interpretation of the results. Quantitative estimates of these effects are carried out via the study of proton--nucleus collisions~\cite{Abelev:2013yxa,Acharya:2018kxc,Sirunyan:2017mzd,Adam:2015iga,Aaij:2017cqq,Aaboud:2017cif}.
Finally, a quantitative interpretation of the results requires taking into account that the observed \jpsi\ are produced either promptly, i.e. as direct \jpsi\ or via decay of higher-mass charmonium states ($\chi_{\rm c}$, $\psi(2S)$), or non-promptly through the weak decay of hadrons containing a b quark~\cite{Andronic:2015wma}.

For a better assessment of the suppression-regeneration scenario, extensive studies of the centrality, \pt\ and rapidity dependence of the \jpsi\ nuclear modification factor have to be carried out. The first \mbox{ALICE} measurement of the inclusive (sum of prompt and non-prompt sources) \jpsi\ production at \sqrtsnnE{5.02} at forward rapidity~\cite{Adam:2016rdg} has shown a hint for an increase of \Raa\ with respect to the \sqrtsnnE{2.76} results in the region 2 $<$ \pt\ $<$ 6 \gevc, while the results were consistent elsewhere.
%small increase of \Raa, both as a function of centrality and \pt, with respect to the \sqrtsnnE{2.76} results, although the measurements at the two energies are compatible within uncertainties

In this paper, we complement the results obtained in Ref.~\cite{Adam:2016rdg}. The \jpsi\ \Raa\ is simultaneously obtained in different collision centrality classes and \pt\ or rapidity intervals. In addition, to further assess the kinematic region of influence of the \jpsi\ regeneration mechanism, results on the \jpsi\ average \pt\ and $\pt^2$ as a function of centrality are presented. 

The paper is organized as follows: Sec.~\ref{sec:detector} is dedicated to the description of the \mbox{ALICE} detector systems used in this analysis. The analysis procedure is briefly explained and a summary of the systematic uncertainties is also given in Sec.~\ref{sec:analysis}. Results are presented and compared to available measurements at \sqrtsnnE{2.76} and model calculations in Sec.~\ref{sec:results}.

%This observation has a natural explanation if a further \jpsi\ production mechanism, known as regeneration, predicted to occur in high-energy collisions and related to the recombination~\cite{Thews:2000rj,BraunMunzinger:2000px} of c$\overline{\rm c}$ pairs, abundantly produced in such collisions, is considered. 
%
%__________________________________________________________
\section{Apparatus and data sample}
\label{sec:detector}
The ALICE detector and its performance are extensively described in Refs.~\cite{ALICE1} and~\cite{ALICE2}, respectively. \jpsi\ mesons are reconstructed in the muon spectrometer (covering the pseudo-rapidity interval $-4$ $<$ $\eta$ $<$ $-2.5$\footnote{In the ALICE reference frame, the muon spectrometer covers a negative $\eta$ interval and consequently negative \y\ values. We have chosen to present our results with a positive \y\ notation.}) via their dimuon decay channel down to zero \pt. The muon spectrometer consists of a \mbox{4.1 m} (10 interaction lengths ($\lambda_{\rm int}$)) thick front absorber which is used to filter out hadrons coming from the interaction point (IP), followed by tracking (MCH) and triggering (MTR) systems. Each of the five tracking stations is composed of two planes of cathode pad chambers. The third tracking station is located inside a dipole magnet with a field integral of 3 Tm. A 1.2 m (7.2$\lambda_{\rm int}$) thick iron wall, which absorbs secondary hadrons escaping from the front absorber and low-momentum muons produced predominantly from $\pi$ and K decays, is located between the tracking system and the trigger stations. Each of the two trigger stations consists of two planes of resistive plate chambers. Finally, a small-angle conical absorber around the beam-pipe protects the spectrometer from secondary particles produced by interactions of large-$\eta$ primary particles with the beam-pipe.

The other detectors used in this analysis are the Silicon Pixel Detector (SPD), the V0 scintillator detectors, the Cherenkov detectors T0 and the Zero Degree Calorimeters (ZDC). The SPD~\cite{Aamodt:2010aa} provides the coordinates of the primary vertex of the collision, and consists of two cylindrical layers covering $|\eta|$ $<$ 2 (inner layer) and $|\eta|$ $<$ 1.4 (outer layer). The V0~\cite{Abbas:2013taa}, composed of two arrays of 32 scintillator tiles each, and located on both sides of the IP, covers 2.8 $<$ $\eta$ $<$ 5.1 (V0A) and $-3.7 < \eta < -1.7$ (V0C), and is used as a trigger detector, for the centrality determination and to remove beam-induced background. It is also used for the measurement of the luminosity along with the T0 detector~\cite{1546489}, which consists of two quartz Cherenkov counters, located on each side of the IP and covering the pseudo-rapidity intervals $-3.3< \eta < -3$ and $4.6 < \eta < 4.9$. The ZDCs, located on either side of the IP at $\pm$ 114 m along the beam axis, detect spectator nucleons emitted at zero degrees with respect to the LHC beam axis, and are used to reject electromagnetic \PbPb\ interactions~\cite{ALICE:2012aa}.

The centrality determination and the evaluation of the average number of participant nucleons in the collision ($\langle N_{\rm part}\rangle$) for each centrality class is based on a Glauber model fit to the V0 signal amplitude distribution as described in Refs.~\cite{Cent1,Cent2}. The events are classified in centrality classes corresponding to percentiles of the nuclear hadronic cross section. In this analysis, events corresponding to the most central 90\% of the inelastic cross section were selected. For these events the minimum bias (MB) trigger is fully efficient and the residual contamination from electromagnetic processes is negligible. The MB trigger is defined by a coincidence of the signals from both sides of the V0 detector.% The systematic uncertainty on the definition of the centrality intervals is evaluated by varying by $\pm 0.5$\% the value of the V0 signal amplitude corresponding to 90\% centrality and redefining accordingly the centrality intervals, following the approach detailed in~\cite{Adam:2016rdg}.

%no details on pp

The analysis presented here is based on dimuon-triggered events which require, in addition to the MB condition, the detection of two Unlike-Sign (US) tracks in the triggering  system of the muon spectrometer. The muon trigger selects muon candidates having a transverse momentum larger than a given threshold which corresponds to the value for which the trigger efficiency reaches 50\%. In \PbPb\ collisions the \pt\ threshold is $\approx$ 1 \gevc\ with the single-muon trigger efficiency reaching a plateau value of 98\% at $\approx$ 2.5 \gevc~\cite{Collaboration_2012}.

The current analysis exploits the data samples of \PbPb\ collisions at \sqrtsnn\ $=$ 5.02 TeV collected during 2015. This corresponds to an integrated luminosity $L^{\rm PbPb}_{\rm int}$ $\approx$ 225 $\mu$b$^{-1}$.

% for \PbPb\ collisions and $L^{\rm pp}_{int}$ $\approx$ 106 nb$^{-1}$ for pp collisions~\cite{Acharya:2017hjh}. 

%Like-sign dimuon triggers were also collected, used mainly for background normalisation purposes in the Pb-Pb analysis.
%
%The integrated luminosity corresponding to the analysed data samples are $L^{\rm Pb-Pb}_{int}$ $\approx$ 225 $\mu$b$^{-1}$ for Pb-Pb~\cite{Adam:2016rdg} and $L^{\rm pp}_{int}$ $\approx$ 106 nb$^{-1}$ for pp~\cite{Acharya:2017hjh} collisions.

\section{Data analysis}
\label{sec:analysis}
%The analysis procedure followed in this paper is the same adopted in~\cite{Adam:2015isa}
%Below, the definitions of the measured observables are given, followed by a presentation of the main analysis steps and a discussion of the systematic uncertainties. 
For a centrality class $i$, the double-differential \jpsi\ invariant yield ($Y^i_{\jpsi}$) is defined as
 \begin{eqnarray}
		 \frac{{\rm d}^2Y^i_{\jpsi}}{{\rm d}y{\rm d}\pt} = \frac{N_{{\rm J}/\psi}^{i}(\pt,\y)}{{\rm BR}_{\rm{J/}\psi\rightarrow\mu^{+}\mu^{-}} \cdot \Delta \pt \cdot \Delta y \cdot (\Ae)^{i}(\pt,\y) \cdot N_{\rm{MB}}^{i}},
	\label{eq:yield}
	\end{eqnarray}
 
% where:
%	\begin{itemize}
%		\item $N_{{\rm J}/\psi}^{i}(\Delta \pt,\Delta y)$ is the number of \jpsi\ for a given \pt\ and \y\ interval,
%		\item ${\rm BR}_{\rm{J/}\psi\rightarrow\mu^{+}\mu^{-}} = (5.96\pm 0.03)\%$ is the branching ratio of the dimuon decay channel~\cite{Tanabashi:2018oca},
%		\item $\Ae^{i}(\Delta \pt,\Delta y)$ is the product of the detector acceptance and the reconstruction efficiency for a given \pt\ and \y\ interval,
%		\item $N_{\rm{MB}}^{i}$ is the equivalent number of minimum-bias events. It is obtained as the product of the number of dimuon-triggered events times the inverse of the probability of having a dimuon trigger in a MB event ($F^{i}$). The $F^{i}$ values correspond to those quoted in~\cite{Adam:2016rdg}. For the centrality integrated sample the numerical value of the normalization factor is $F^{0-90\%}=11.84\pm 0.06$. The quoted uncertainty is systematic and corresponds to the difference between the results obtained with two methods, either by calculating the ratio of the counting rates of the two triggers, or by applying the dimuon trigger condition in the analysis of MB events.
%	\end{itemize}

where $N_{{\rm J}/\psi}^{i}(\pt,\y)$ is the number of \jpsi\ for a given \pt\ and \y\ interval, ${\rm BR}_{\rm{J/}\psi\rightarrow\mu^{+}\mu^{-}} = (5.96\pm 0.03)\%$ is the branching ratio of the dimuon decay channel~\cite{Tanabashi:2018oca}, $\Delta \pt$ and $\Delta y$ are respectively the widths of the \pt\ and \y\ intervals, $(\Ae)^{i}(\pt,\y)$ is the product of the detector acceptance and the reconstruction efficiency for that \pt\ and \y\ interval, and $N_{\rm{MB}}^{i}$ is the equivalent number of minimum-bias events. The values of $N_{\rm{MB}}^{i}$ are obtained as the product of the number of dimuon-triggered events times the inverse of the probability of having a dimuon trigger in a MB event ($F^{i}$). The $F^{i}$ values correspond to those quoted in Ref.~\cite{Adam:2016rdg}. For the centrality integrated sample, the value of the normalization factor is $F^{0-90\%}=11.84\pm 0.06$. The quoted uncertainty is systematic and corresponds to the difference between the results obtained with two methods, either by calculating the ratio of the counting rates of the two triggers, or by applying the dimuon trigger condition in the analysis of MB events.\\
	The nuclear modification factor \Raa\ is given by

	\begin{eqnarray}
		R_{\rm{AA}}^{i}(\pt,\y) = \frac{{\rm d}^2Y^i_{\jpsi}/{\rm d}y{\rm d}\pt}{\Taa^{i} \cdot {\rm d}^2 \sigma_{\rm{J/}\psi}^{\rm{pp}}/{\rm d}y{\rm d}\pt},
	\label{eq:Raa}
	\end{eqnarray}
	
	where $\Taa^{i}$ is the average of the nuclear-overlap function~\cite{Miller:2007ri}. The values of $\Taa^{i}$ in different centrality classes were obtained using a Glauber calculation~\cite{Cent2,ALICE-PUBLIC-2018-011,Loizides:2017ack}. The systematic uncertainty on the $\Taa^{i}$ calculation, which ranges from $1\%$ in the most central class to $3\%$ in the most peripheral one, was determined by varying the density parameters of the Pb nucleus and the nucleon--nucleon inelastic cross section within their uncertainties. The systematic uncertainty on the definition of the centrality intervals is evaluated by varying by $\pm 0.5$\% the fraction (90\%) of the hadronic cross section selected with the chosen minimal cut on the V0 signal amplitude, and redefining accordingly the centrality intervals, following the approach detailed in Ref.~\cite{Adam:2016rdg}.
Values of the \jpsi\ cross section in \pp\ collisions (${\rm d}^2 \sigma_{\rm{J/}\psi}^{\rm{pp}}/{\rm d}y{\rm d}\pt$) at \sqrtsE{5.02} were already reported in Refs.~\cite{Adam:2016rdg,Acharya:2017hjh} and are used here as a reference. In addition, following the same analysis procedure as detailed in those papers, the cross section was evaluated in four \pt\ intervals (0.3--2, 2--5, 5--8, and 8--12 \gevc) for the interval 2.5 $<$ \y\ $<$ 4 and three \y\ intervals (2.5--3, 3--3.5, and 3.5--4) for the interval \pt\ $<$ 12 \gevc. The integrated luminosity of the \pp\ sample is $L^{\rm pp}_{\rm int} = (106.3 \pm 2.2$) nb$^{-1}$~\cite{ALICE-PUBLIC-2016-005}.

The \jpsi\ candidates were formed by combining US muons reconstructed within the geometrical acceptance of the muon spectrometer using the tracking algorithm described in Ref.~\cite{Aamodt:2011gj}. The selection criteria applied to both single muons and dimuons are identical to the ones used in Refs.~\cite{Adam:2016rdg,Adam:2015isa}, requiring a match between tracks reconstructed in the tracking system and track segments in the muon trigger system.
	
	The signal extraction was performed over the US dimuon invariant mass ranges [2.2,4.5] and [2.4,4.7] \gevcc\ using two methods. In the first one, the invariant mass distributions were fitted with the sum of a signal and a background function. In the second one, the combinatorial background (dominant in central \PbPb\ collisions) was first estimated using an event-mixing technique~\cite{Adam:2015isa}, and then subtracted from the raw invariant mass distribution. Finally, the resulting distributions were fitted with the sum of a signal and a residual background component.
%Note that the second method was not used to extract the \jpsi\ signal in \pp\ collisions.
	
	The signal component of the fitting function is either a double-sided Crystal Ball function (CB2, where independent non-Gaussian tails are present on both sides of a Gaussian core) or a pseudo-Gaussian with a mass-dependent width~\cite{ALICE-PUBLIC-2015-006}. For both functions, the position of the \jpsi\ pole mass, as well as the width of the resonance, are free parameters of the fits, while the non-Gaussian tail parameters were fixed.
%from Monte Carlo (MC) simulations. For the \PbPb\ analysis, $\jpsi\rightarrow \mumu $ signals are embedded into real events to take into account the detector occupancy.
%Two sets of tail parameters were obtained from Monte Carlo (MC) simulations using  different particle transport codes (GEANT3~\cite{Brun:1994aa} and GEANT4~\cite{Agostinelli:2003250}) to account for the sensitivity of these  parameters to the description of the detector materials. In addition, another set of CB2 tail parameters was extracted from the \pp\ collisions at \sqrtsE{13} data sample~\cite{Acharya:2017hjh}, the sample with the largest significance of the \jpsi\ signal. Finally, the contribution of the \psip\ signal was also taken into account in the fits, using the same signal functions as for the \jpsi, with mass and width tied to those of the \jpsi~\cite{Abelev:2014zpa}.
%	
Two sets of tail parameters were obtained from Monte Carlo (MC) simulations using  different particle transport codes (GEANT3~\cite{Brun:1994aa} and GEANT4~\cite{Agostinelli:2003250}) to account for the sensitivity of these  parameters to the description of the detector materials. In addition, another set of CB2 tail parameters was extracted from the \pp\ collisions at \sqrtsE{13} data sample~\cite{Acharya:2017hjh}, the sample with the largest significance of the \jpsi\ signal. The \psip\ signal was included in the fits using the same signal functions as for the \jpsi, with mass and width tied to those of the \jpsi~\cite{Abelev:2014zpa}.
	
\begin{figure}[th!]
			\centering
			\includegraphics[width=0.95\textwidth]{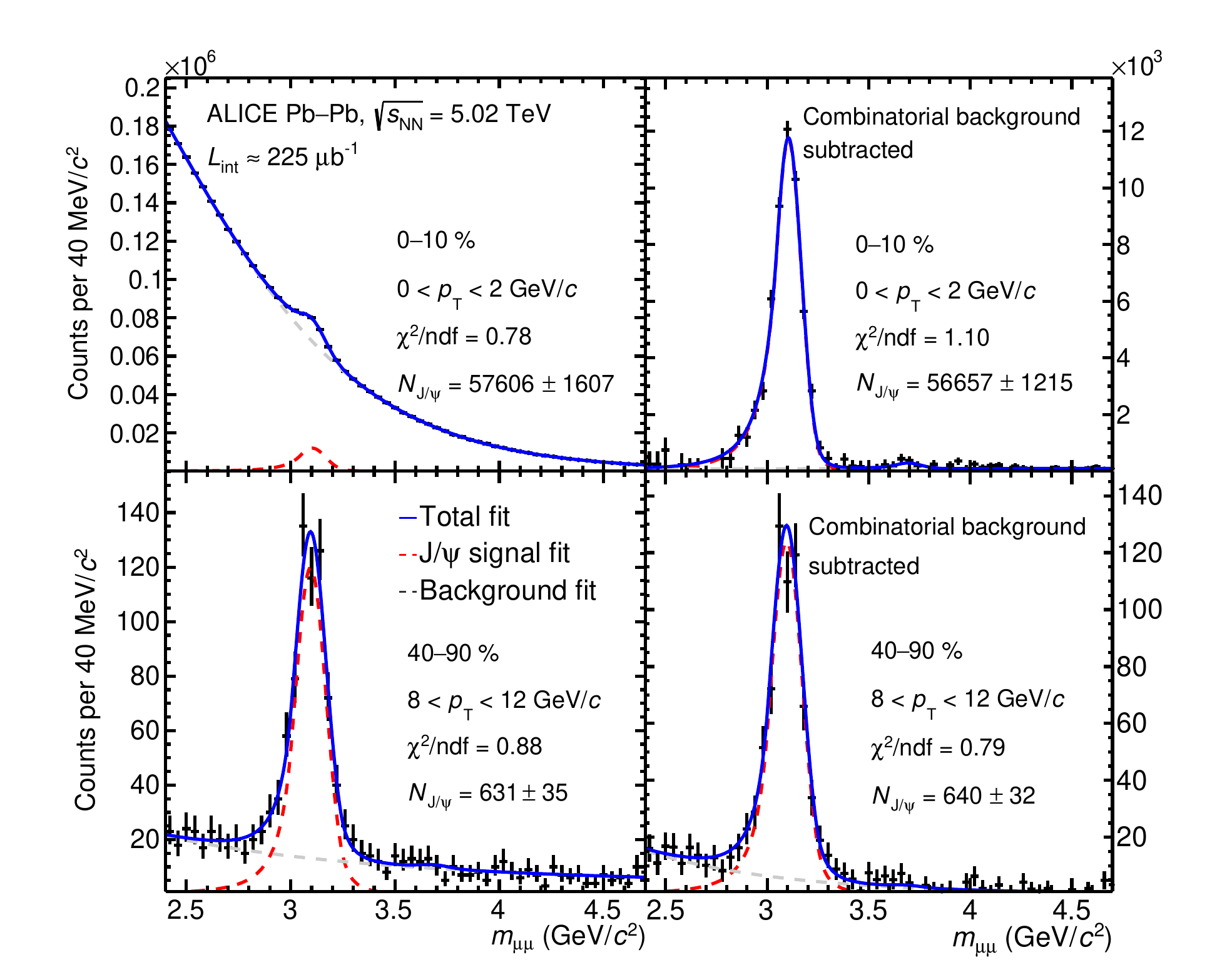}
			\caption{Example of fits to the US dimuon invariant mass distributions in \PbPb\ collisions at \sqrtsnnE{5.02} in different centrality classes and \pt\ intervals. The left (right) panels show the distributions before (after) background subtraction with the event-mixing technique. Dashed gray curves correspond to background functions, red curves to the signal functions and blue curves to the sum of the signal and background functions.}
			\label{fig:InvMassSpec}
		\end{figure}

	The background was parametrized either as a pseudo-Gaussian with a width quadratically dependent on the mass or as a ratio of a 2\textsuperscript{nd} to 3\textsuperscript{rd} order polynomial. When using the event-mixing technique, the continuum component of the correlated background remaining in the US dimuon distributions after the background subtraction and originating mainly from semi-muonic decays of pairs of charm hadrons, was parametrized using the sum of two exponential functions. Examples of fits to the US dimuon invariant mass distributions, without and with subtraction of mixed-event background, are shown in Fig.~\ref{fig:InvMassSpec} for different centrality classes and \pt\ intervals. For each centrality class, \pt\ and \y\ interval, the number of \jpsi\ and the statistical uncertainty are given by the average of the results from the considered fit configurations obtained by varying the signal and background functions, the tail parameters and the invariant mass fit range. The systematic uncertainty is defined, for each centrality, \pt\ and \y\ interval, as the RMS of the various fit results. It varies between 1.5\% and 3.6\% as a function of centrality or \pt\ and between 1.5\% and 5\% as a function of \y.
%	
%	It varies between 1.5\% and 3.6\% for the various centrality/\pt\ ranges integrated in \y, and between 1.5\% and 5.0\% versus \y\ in the different centrality intervals.

	%_____________________________
%\subsection{Acceptance and efficiency correction}
The \jpsi\ \Ae\ was obtained using MC simulations, where the \pt\ and $y$ distributions for the generated \jpsi\ were matched to the ones extracted from data using an iterative procedure as done in Ref.~\cite{Acharya:2018kxc}. Unpolarized J/$\psi$ production was assumed, consistently with the measurements of inclusive \jpsi\ polarization in \pp\ collisions~\cite{Abelev:2011md,Acharya:2018uww}. The misalignment of the detection elements as well as the time-dependent status of each electronic channel during the data taking period were taken into account in the simulation. Generated $\jpsi \rightarrow \mu^{+}\mu^{-}$ signals were embedded into real minimum bias events in order to properly reproduce the effect of detector occupancy and its variation from one centrality class to another, and reconstructed as for real events. A relative decrease by $\sim$14\%  of \Ae\ was observed in the most central \PbPb\ collisions with respect to the most peripheral ones. 
%Finally, the \Ae\ was computed as the ratio between the number of \jpsi\ reconstructed in the muon spectrometer and the number of generated ones in the same centrality, \pt, \y\ intervals. 
 
 % Furthermore, they were re-weighted to reproduce the data in the different centrality classes.
% In the \PbPb\ analysis, additional weights were applied for these distributions to take into account the variation of the \pt\ and $y$ shapes in various centrality intervals.
% In the \PbPb\ analysis, the adjustment is done using centrality integrated (0--90\%) data  and a separate weighting for each centrality interval is done based on the \pt\ and $y$ distributions for that centrality. 

 %Track reconstruction and signal extraction are performed from the simulated hits generated in the detector using the same procedure and selection criteria as those used for the data.
	
	The following sources of systematic uncertainty on \Ae\ were considered: \begin{enumerate*}[label=(\roman*)]
		\item the parametrization of the input $\pt$ and $\y$ shapes,
			%and the correlation between these,
		\item the uncertainty on the tracking efficiency in the muon tracking chambers,
		\item the uncertainty on the MTR efficiency and
		\item the matching between tracks reconstructed in the tracking and triggering systems.
	\end{enumerate*}
	
For the parametrization of the MC input distributions, two sources of systematic uncertainty were considered: the effect of the finite data sample used to parametrize these distributions and the correlations between \pt\ and \y\ (more explicitly, the fact that the \pt\ distribution of the \jpsi\ varies within the rapidity interval in which it is measured). The former turns out to be negligible. For the latter, different MC simulations were performed by varying the input \pt\ and \y\ distributions within limits that correspond to this effect and re-calculating the \Ae\ in each case as done in Ref.~\cite{Acharya:2017hjh}. The uncertainties on the tracking efficiency in the MCH, trigger efficiency in the MTR, and on the matching efficiency between MTR and MCH tracks were evaluated by comparing the efficiencies obtained in data and MC at the single muon level and propagating the observed differences to the \jpsi\ candidates, as done in Ref.~\cite{Abelev:2014qha}.

     In each centrality, $\pt$ and $y$ interval, the total systematic uncertainty on the yield and \Raa\ is determined as the quadratic sum of the uncertainties from the different sources listed in Table~\ref{tab:syst_PbPb}. Correlations of various uncertainties vs centrality, \pt\ or \y\ are also reported. The values in the last row correspond to the sum of the statistical and systematic uncertainties on the \pp\ reference.

%     Table~\ref{tab:syst_pp} summarizes the different sources of systematic uncertainties on $\sigma_{\rm{J/}\psi}$ in \pp\ collisions at \sqrts\ = 5.02 TeV in \pt\ and \y\ intervals.
     % that are not reported in Ref.~\cite{Acharya:2017hjh}.
     
     \begin{table}[th!]
	\centering
	\caption{Summary of systematic uncertainties, in percentage, on the yield and $R_{\rm AA}$ in \PbPb\ collisions at \sqrtsnnE{5.02}. Values with an asterisk correspond to the systematic uncertainties correlated as a function of the given variable. For the pp reference, the correlated and uncorrelated contributions are separated.}
	\vspace{5mm}
	\begin{tabular}{l|c|c|c}
    \hline
    Sources        &vs centrality&vs \pt\ & vs \y\ \\
    \hline
    Signal extraction  & 1.5--3.6 & 1.5--3.6& 1.5--5.0 \\

    MC input      &   2$^*$	& 2--3		& 0.5--2.5  \\
    
    Tracking efficiency & 3$^*$ & 3  & 3  \\
    
    Trigger efficiency  & 1.5--2.7 $^*$ & 1.5--4.1 & 1.5--2.4 \\
    
    Matching efficiency & 1$^*$ & 1  & 1 \\
    
    $F$ & 0.5$^*$  &  0.5$^*$ & 0.5$^*$\\
    
    BR (only on yield) \cite{Tanabashi:2018oca} & 0.5$^*$  &  0.5$^*$ & 0.5$^*$\\
    
    $\langle T_{\rm AA}\rangle$ (only on \Raa)&   0.7--3.2 & 0.7--2.0$^*$ & 0.7--2.0$^*$\\
    
    Centrality definition&  0.1--3.5  & 0.2--1.4$^*$ & 0.2--1.4$^*$ \\
    
    pp reference (only on \Raa)& 4.9--10.9$^*$ & 4.4--16.5 and 2.1$^*$ & 4.7--8.5 and 2.1$^*$ \\
    \hline
  \end{tabular}
  \label{tab:syst_PbPb}
\end{table}

\section{Results}
\label{sec:results}

\subsection{Nuclear modification factor}
\label{sec:Raa}

	This section summarizes the results for the inclusive \jpsi\ \Raa\  at forward rapidity in \PbPb\ collisions at \sqrtsnnE{5.02} as a function of:

	\begin{itemize}
		\item  rapidity and transverse momentum, integrated over the centrality (class 0--90\%);
		\item  rapidity and transverse momentum, for the centrality classes 0--20\%, 20--40\% and 40--90\%;
		\item  centrality, in four transverse momentum intervals and in three rapidity intervals.
	\end{itemize}

	When possible, the ratio between the results of this analysis and the measurements in \PbPb\ collisions at \sqrtsnnE{2.76}~\cite{Adam:2015isa}, in the same kinematic interval, is computed. Only the uncertainty related to the $\langle T_{\rm AA}\rangle$ cancels out in the ratio, as discussed in Ref.~\cite{Adam:2016rdg}. Following the same approach as in Refs.~\cite{Adam:2016rdg,Adam:2015isa}, an estimate of the \Raa\ of prompt \jpsi\ was determined by making conservative assumptions on the size of the non-prompt \Raa. The relation between the inclusive (\Raa), prompt ($\Raa^{\rm prompt}$) and non-prompt ($\Raa^{ \rm non\textnormal{-}prompt}$) nuclear modification factors can be expressed as:
	
\begin{equation}\label{eq:prompt}
\Raa^{\rm prompt} = \frac{\Raa - F_{\rm B} \cdot \Raa^{\rm non\textnormal{-}prompt}}{1-F_{\rm B}},
\end{equation}

where $F_{\rm B}$ is the fraction of non-prompt to inclusive \jpsi\ in pp collisions. This quantity is evaluated at \sqrtsE{5.02} by interpolating in energy the corresponding LHCb cross-section measurements in pp collisions at $\sqrt{s}=$ 2.76 and 7 TeV~\cite{Aaij:2011jh,Aaij:2012asz,LHCb-CONF-2013-013}. The limits on $\Raa^{\rm prompt}$ correspond to the two extreme hypotheses of total non-prompt \jpsi\ suppression ($\Raa^{\rm non\textnormal{-}prompt}=0$) and absence of suppression ($\Raa^{\rm non\textnormal{-}prompt}=1$). The effect is small at moderate transverse momentum ($\lesssim$ 10\% for \pt\ $\lesssim$ 5 \gevc) and then increases at higher \pt. Numerical values for the limits on $\Raa^{\rm prompt}$ can be found in the HepData record associated to this paper. Another effect which may influence the interpretation of the inclusive \jpsi\ results is the presence of an excess at very low \pt, observed at \sqrtsnnE{2.76}~\cite{Adam:2015gba} and related to \jpsi\ photo-production~\cite{Abelev:2012ba}. This source was shown to be significant with respect to hadronic production for peripheral \PbPb\ collisions and has a strong influence on the measured \Raa\ values. For this reason, the region \pt\ $<$ 0.3 \gevc\ was excluded when dealing with peripheral collisions. The remaining contribution of this source to the region \pt\ $>$ 0.3 \gevc\ was evaluated following the procedure detailed in Ref.~\cite{Adam:2015isa} and the maximum effect on \Raa\ is explicitly shown in the following figures by use of bracket symbols. The upper and lower limit brackets correspond to the extremest hypotheses on the contribution from photo-produced \jpsi\ and on the efficiency of the aforementioned \pt\ selection as described in Ref.~\cite{Adam:2015isa}.

	\subsubsection{Centrality-integrated \Raa\ as a function of \y\ and \pt}
\begin{figure}[!htb]
\begin{center}
\includegraphics[width=0.65\textwidth]{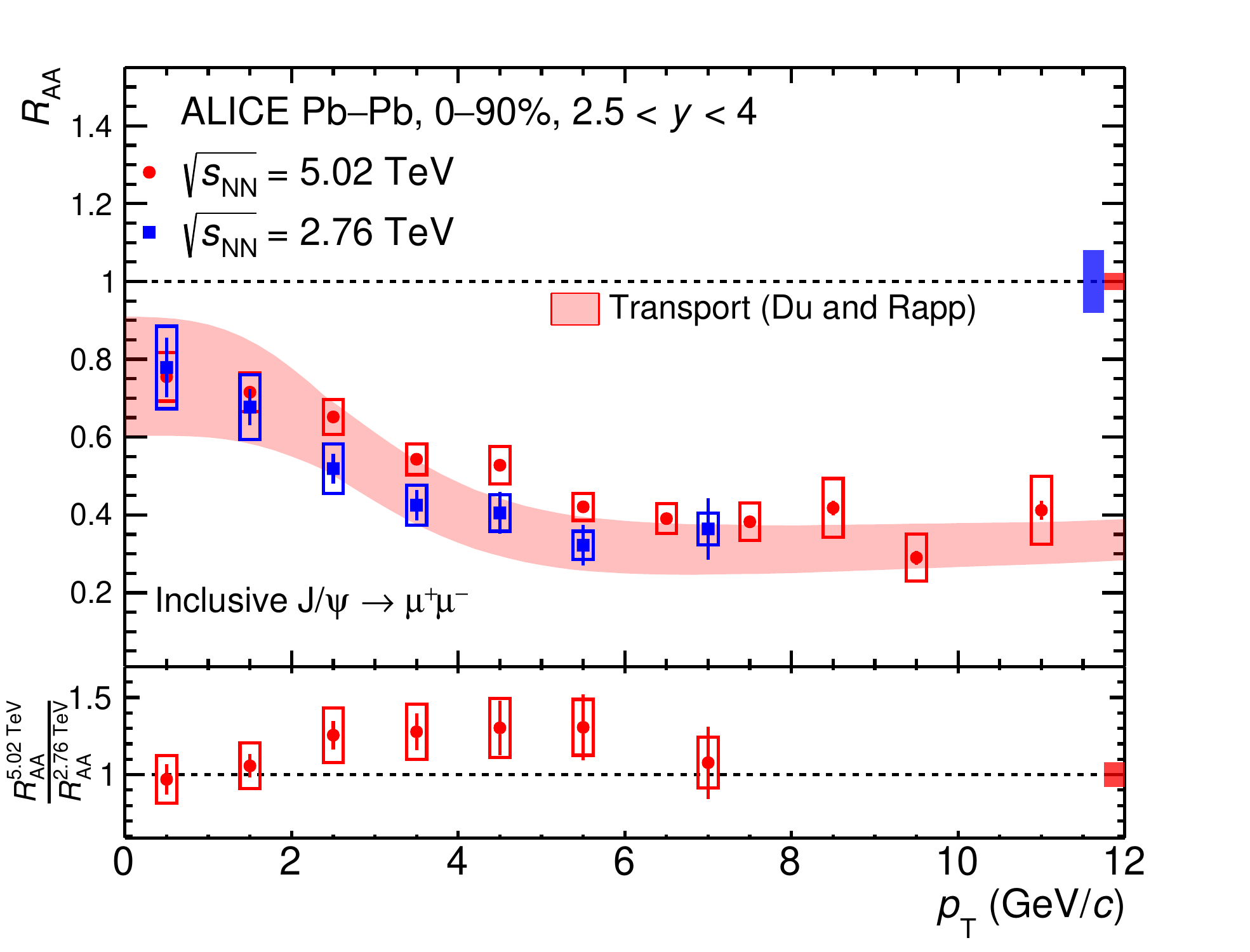}
\caption{Inclusive \jpsi\ nuclear modification factor as a function of \pt\ for \PbPb\ collisions at \sqrtsnnE{5.02} in the 0--90\% centrality class. The vertical error bars represent statistical uncertainties, the boxes around the points uncorrelated systematic uncertainties, while the correlated uncertainty is shown as a filled box around $\Raa=1$. The corresponding measurements in \PbPb\ collisions at \sqrtsnnE{2.76}~\cite{Adam:2015isa} are also shown, as well as the ratio of the $R_{\rm AA}$ values, which is depicted in the bottom panel of the figure. The \Raa\ values at \sqrtsnnE{5.02} are compared with transport model calculations~\cite{Du:2015wha}.}
\label{fig:2}
\end{center}
\end{figure}

\begin{figure}[!htb]
\begin{center}
\includegraphics[width=0.65\textwidth]{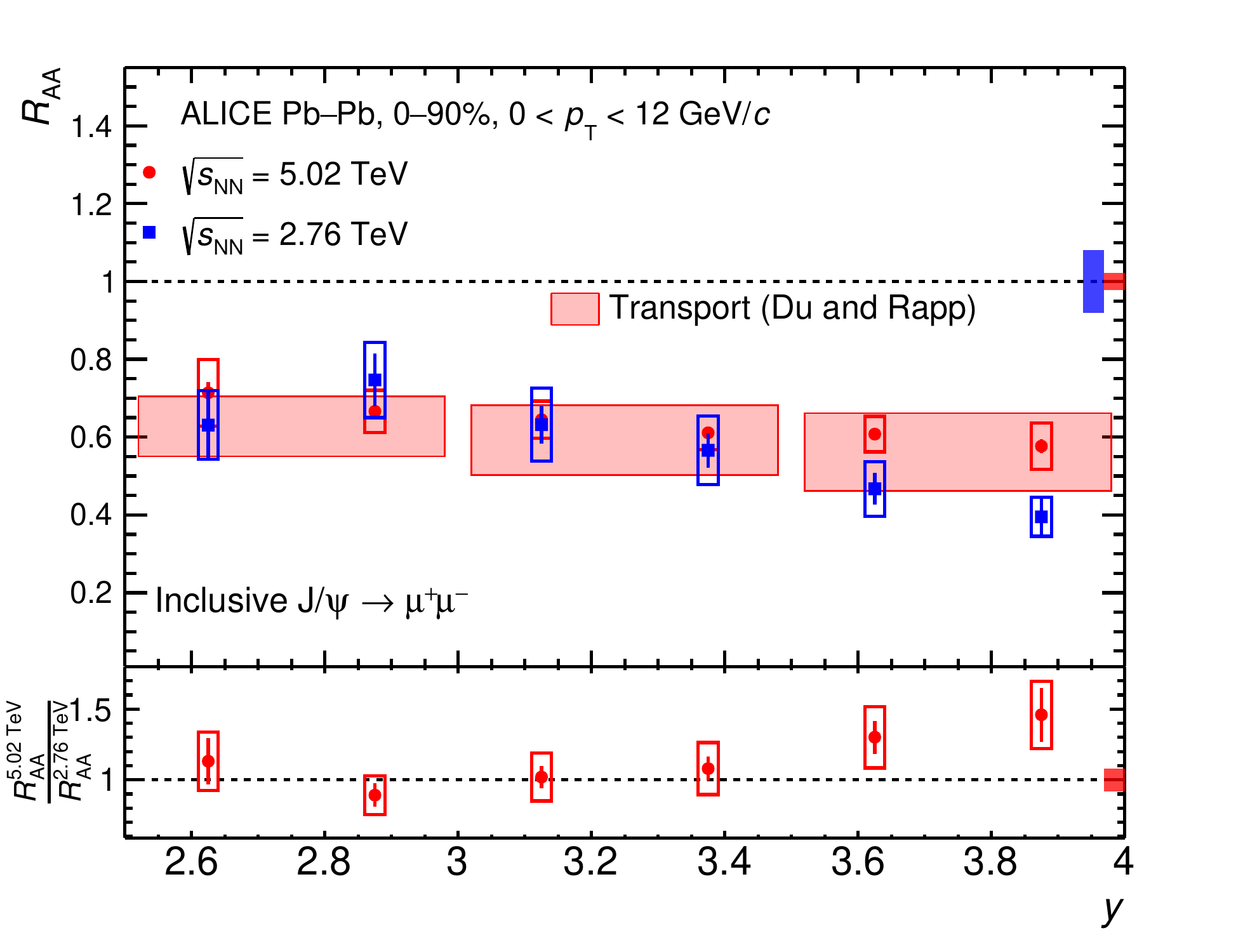}
\caption{Inclusive \jpsi\ nuclear modification factor as a function of rapidity for \PbPb\ collisions at \sqrtsnnE{5.02} in the 0--90\% centrality class. The vertical error bars represent statistical uncertainties, the boxes around the points uncorrelated systematic uncertainties, while the correlated uncertainty is shown as a filled box around $\Raa=1$. The corresponding measurements in \PbPb\ collisions at \sqrtsnnE{2.76}~\cite{Adam:2015isa} are also shown, as well as the ratio of the $R_{\rm AA}$ values, which is depicted in the bottom panel of the figure. The \Raa\ values  at \sqrtsnnE{5.02} are compared with transport model calculations~\cite{Du:2015wha}.}
\label{fig:3}
\end{center}
\end{figure}

Figures~\ref{fig:2} and~\ref{fig:3} show the inclusive \jpsi\ nuclear modification factor as a function of transverse momentum and rapidity, integrated over the centrality class 0--90\%. The results are compared with those obtained at \sqrtsnnE{2.76}~\cite{Adam:2015isa} and with the results of the calculation of a transport model~\cite{Zhao:2011cv,Du:2015wha}. A significant increase of $R_{\rm AA}$ is visible with decreasing \pt, which was already observed for the most central events (0--20\%) and reported in Ref.~\cite{Adam:2016rdg}. Within uncertainties, the results are compatible with those obtained, in a more restricted \pt\ interval, at the lower LHC energy, with a possible hint (1.2$\sigma$) of a weaker suppression in the region 2 $<$ \pt\ $<$ 6 \gevc. The transport model calculations are in qualitative agreement with the data. In this model, a competition between suppression and regeneration of charmonia is assumed, choosing a c$\overline{\rm c}$ production cross section ${\rm d}\sigma_{\rm c\overline{c}}/{\rm d}y$ $=$ 0.57 mb and ${\rm d}\sigma^{\rm pp}_{{\rm J}/\psi}/{\rm d}y$ $=$ 3.35 $\mu$b for 2.5 $<$ \y\ $<$ 4. The latter value is $\sim$10\% smaller than our measurement of the same quantity~\cite{Adam:2016rdg}. The model also includes contributions from both prompt and non-prompt \jpsi. The upper (lower) limit of this calculation corresponds to a 10\% (25\%) contribution of nuclear shadowing. 
%As shown in Fig.~\ref{fig:2}, the increase of $R_{\rm AA}$ at low \pt\ is directly connected, according to the model, with the contribution of J/$\psi$ from recombination processes.
%({\it how was the low \pt\ region treated? Does the first bin exclude the region <0.3 GeV/c?})
%confirms previous ALICE results, obtained at \sqrtsnnE{5.02} in the narrower centrality class 0--20\%~\cite{Adam:2016rdg}

Figure~\ref{fig:3} shows that, in the explored rapidity interval, there is no significant variation of the \Raa\ values. The calculations of the transport model are in good agreement with the experimental results. The comparison of the results with those obtained at \sqrtsnnE{2.76}~\cite{Adam:2015isa} hints (1.5$\sigma$) for a weaker suppression at \sqrtsnnE{5.02} at large \y\ (3.75 $<$ \y\ $<$ 4).
%The values are consistent with those observed at \sqrtsnnE{2.76}~\cite{Adam:2015isa}, with a hint for a possibly weaker suppression at large \y.
% except at large $y$, where a stronger suppression might have been observed (x.x$\sigma$ significance for $3.8<y<4$).

	%_____________________________
	\subsubsection{Centrality-differential \Raa\ as a function of \y\ and \pt}

\begin{figure}[!htb]
\begin{center}
\includegraphics[width=0.49\textwidth]{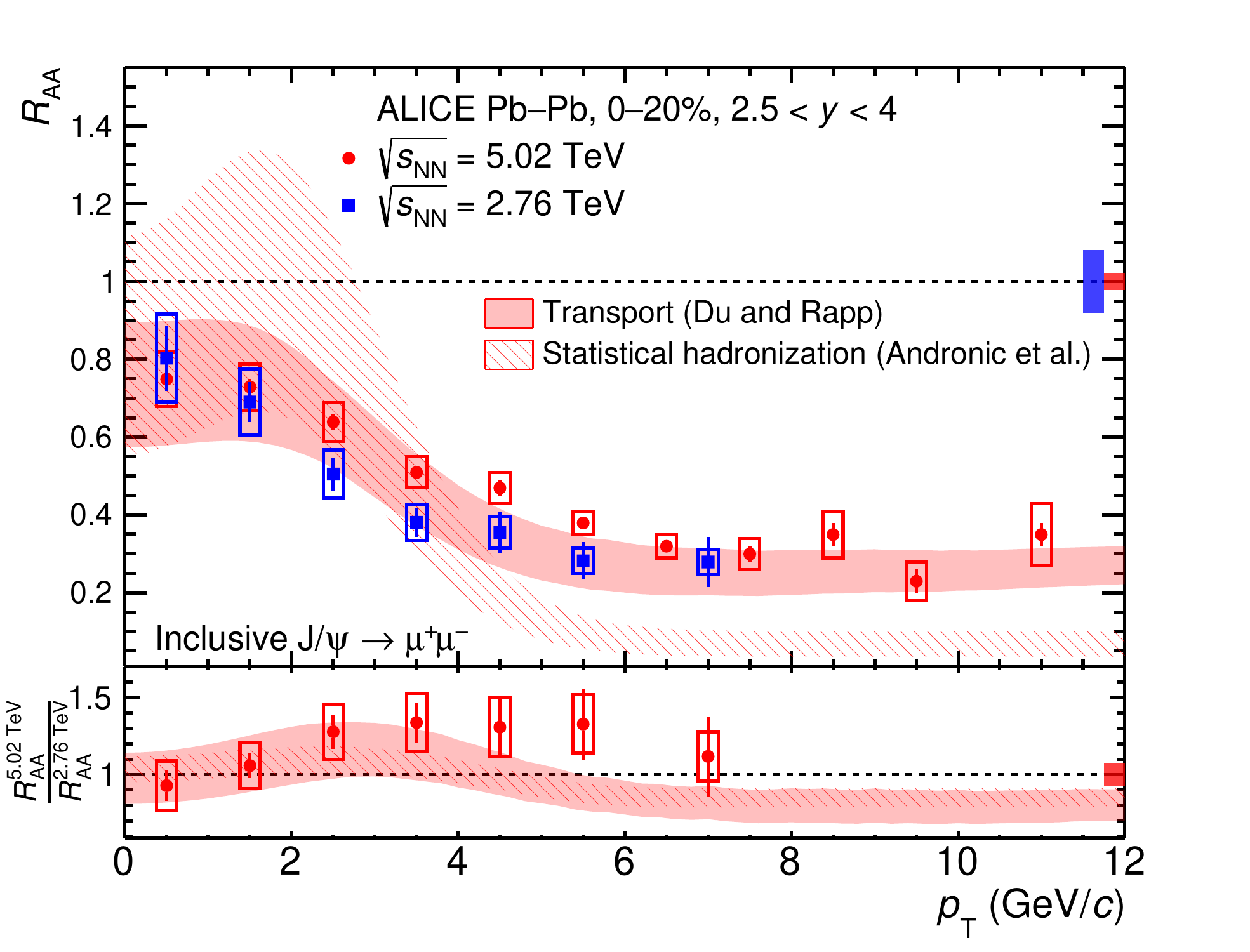}\\
\includegraphics[width=0.49\textwidth]{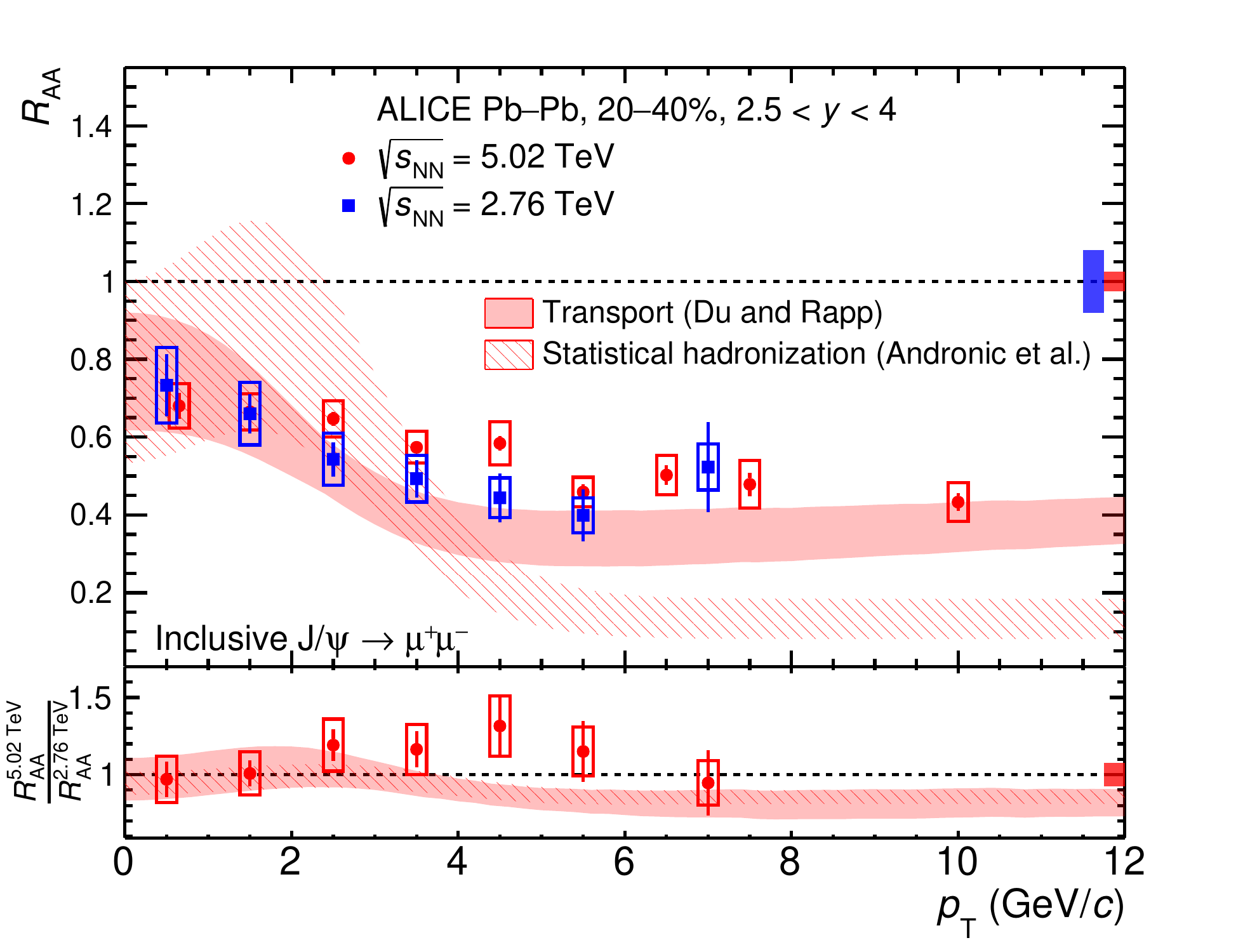}\\
\includegraphics[width=0.49\textwidth]{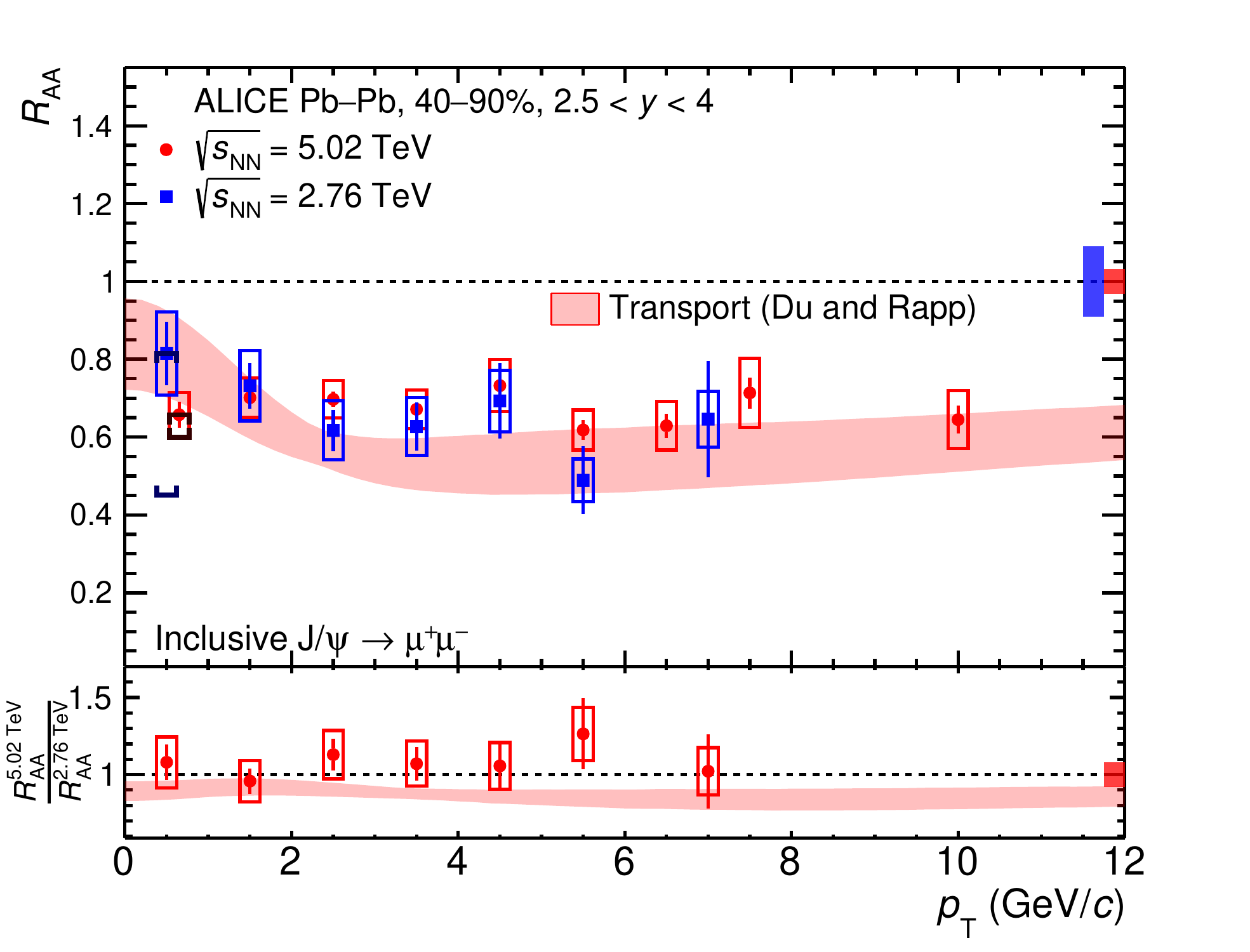}\\
\caption{Inclusive \jpsi\ nuclear modification factor as a function of \pt\ for \PbPb\ collisions at \sqrtsnnE{5.02} in the 0--20\% (top), 20--40\% (middle) and 40--90\% (bottom) centrality classes. The vertical error bars represent statistical uncertainties, the boxes around the points uncorrelated systematic uncertainties, while the correlated uncertainty is shown as a filled box around $\Raa=1$. The corresponding measurements in \PbPb\ collisions at \sqrtsnnE{2.76}~\cite{Adam:2015isa} are also shown, as well as the ratio of the $R_{\rm AA}$ values, which is depicted in the bottom panel of the figure. The \Raa\ values  at \sqrtsnnE{5.02} and the ratios to lower energy results are compared with transport model calculations~\cite{Du:2015wha} and, for 0--20\% and 20--40\% centrality, with the results of the SHM~\cite{Andronic:2019wva}. The brackets around \Raa\ values for 40--90\% centrality in the lowest \pt\ interval represent an estimate of the maximum influence of \jpsi\ photo-production, as detailed in Sec.~\ref{sec:Raa}.}
\label{fig:4}
\end{center}
\end{figure}

\begin{figure}[!htb]
\begin{center}
\includegraphics[width=0.5\textwidth]{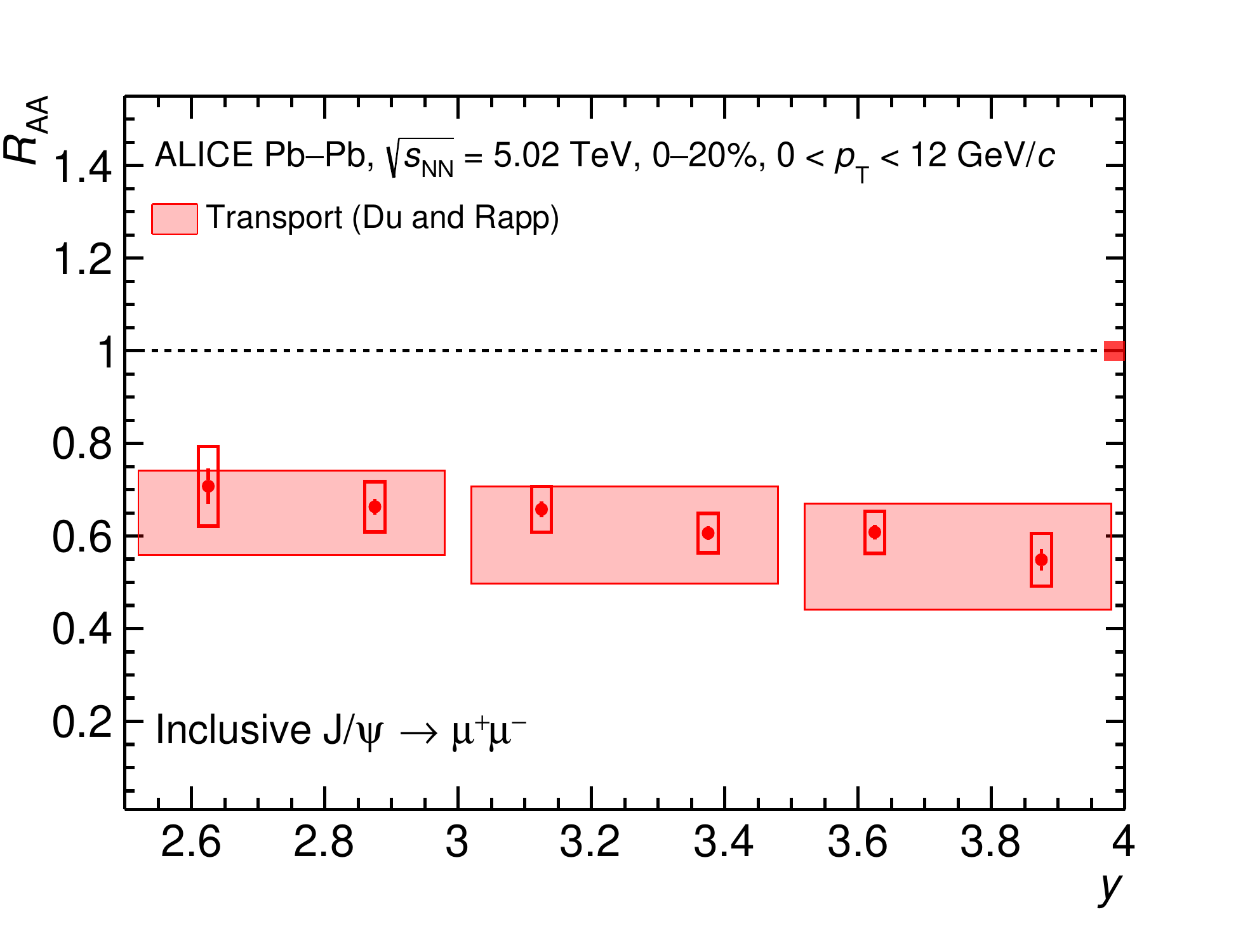}
\includegraphics[width=0.5\textwidth]{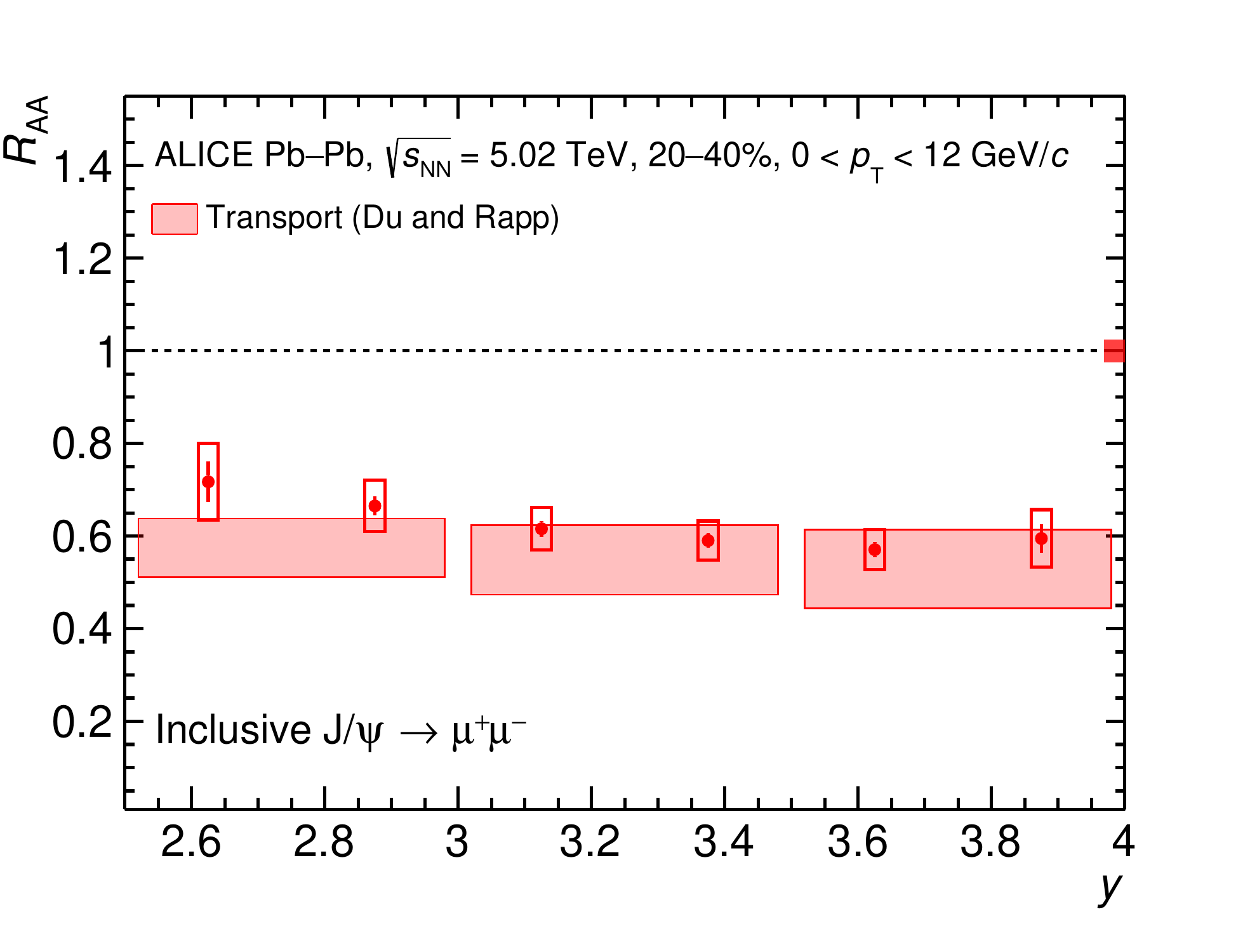}
\includegraphics[width=0.5\textwidth]{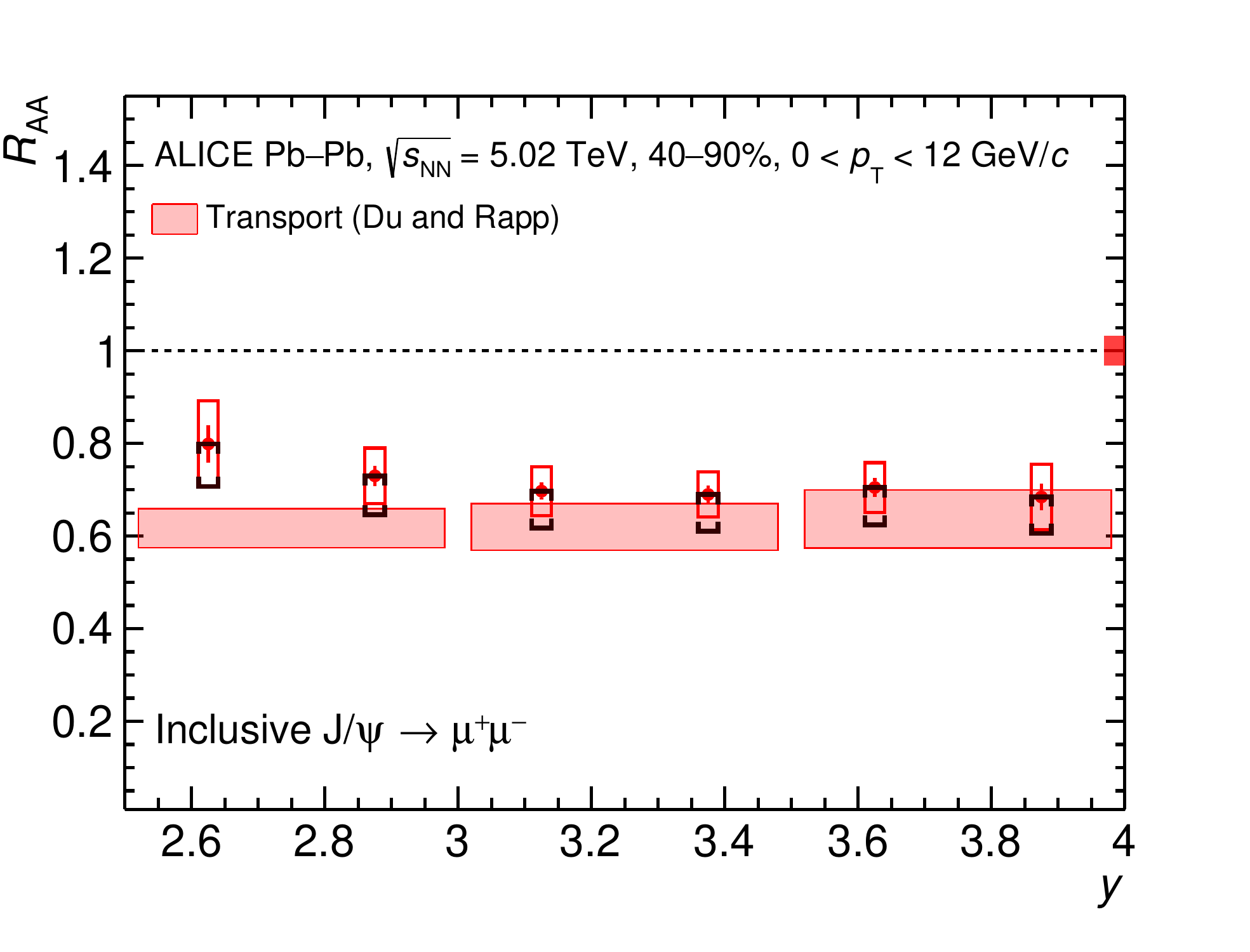}
\caption{Inclusive \jpsi\ nuclear modification factor as a function of rapidity for \PbPb\ collisions at \sqrtsnnE{5.02} in the 0--20\% (top), 20--40\% (middle) and 40--90\% (bottom) centrality classes. The vertical error bars represent statistical uncertainties, the boxes around the points uncorrelated systematic uncertainties, while the correlated uncertainty is shown as a filled box around $\Raa=1$. The \Raa\ values are compared with transport model calculations~\cite{Du:2015wha}. The brackets around \Raa\ values for 40--90\% centrality represent an estimate of the maximum influence of \jpsi\ photo-production, as detailed in Sec.~\ref{sec:Raa}.}
\label{fig:5}
\end{center}
\end{figure}

Figures~\ref{fig:4} and~\ref{fig:5} show, respectively, the \pt\ and $y$ dependence of the inclusive J/$\psi$ $R_{\rm AA}$, for events corresponding to the centrality classes 0--20\%, 20--40\% and 40--90\%. It is worth noting that the results for 0--20\% were already published in Ref.~\cite{Adam:2016rdg}. In this paper, the corresponding values were updated with the improved \Taa\ uncertainties reported in Ref.~\cite{ALICE-PUBLIC-2018-011}.
% and are reported here for completeness. hat the old values were updated reflecting the improved uncertainties on TAA}
%(the results for the most central sample were already published in Ref.~\cite{Adam:2016rdg} and are reported here for completeness).
In Fig.~\ref{fig:4}, moving from central to peripheral collisions, a weaker \pt\ dependence of the \Raa\ is observed, up to an almost constant nuclear modification factor for 40--90\% centrality. When comparing results at \sqrtsnnE{5.02} and 2.76 \tev, a slight increase of the \Raa\ is visible for the most central collisions and for 2 $<$ \pt\ $<$ 6 \gevc\ at the higher collision energy, while the results are compatible in the 20--40\% and 40--90\% samples. A fair agreement with the transport model calculations is observed. The results for the 0--20\% and 20--40\% centrality classes are also compared with a model based on statistical hadronization (SHM)~\cite{Andronic:2019wva}. A good agreement with this calculation, which does not include contributions from non-prompt \jpsi\ production, can be found up to \pt\ $\sim$ 4 \gevc, while at higher transverse momentum \Raa\ is underestimated. This feature could partly be due to additional production mechanisms, not implemented in the model, such as \jpsi\ production from gluon fragmentation in jets.
%, even if the data lie close to the upper edge of the theory uncertainties, in particular at intermediate \pt. 

%The data points in the lowest \pt\ interval for the 20-40\% and 40-90\% centrality classes are slightly shifted, since at \sqrtsnnE{5.02} the analysis was carried out for the restricted range 0.3 $<$ \pt\ $<$ 1 \gevc. 

%As a function of rapidity, in all the centrality intervals, the \Raa\ values shown in Fig.~\ref{fig:5} exhibit the same weak rapidity dependence as observed in 0--90\%
In Fig.~\ref{fig:5}, the \Raa\ values exhibit a very weak rapidity dependence in all the centrality classes, as also observed in 0--90\% (Fig.~\ref{fig:3}). The calculation of the transport model is able to describe the data, in particular when a weak nuclear shadowing scenario (10\%, corresponding to the lower limit chosen by the authors) is adopted.

%\begin{figure}[!htb]
%\begin{center}
%\includegraphics[width=0.7\textwidth]{RAAVsRap_AllCents.pdf}
%\caption{The $y$-dependence of the inclusive J/$\psi$ nuclear modification factor for \PbPb\ collisions at $\sqrt{s_{\rm NN}} = 5.02$ TeV, in the 0--20\% (top), 20--40\% (center) and 40--90\% (bottom) centrality intervals. The vertical error bars represent statistical uncertainties, the boxes around the points uncorrelated systematic uncertainties, while the global uncertainties are shown as a filled box around $\Raa=1$. The \Raa\ values are compared with transport model calculations~\cite{Du:2015wha}.
%The brackets in the lowest \pt\ interval for 40--90\% centrality represent the \Raa\ variation under two extreme hypotheses on the photo-production contribution~\cite{Adam:2015isa}. {\it To be split in three plots for symmetry with Fig.~\ref{fig:4}}.}
%\label{fig:5}
%\end{center}
%\end{figure}

%%%%%Updated: Split into three to be consistent with the pt one

	%_____________________________
	\subsubsection{Centrality dependence of \Raa}
\begin{figure}[htb]
\begin{center}
\includegraphics[width=0.45\textwidth]{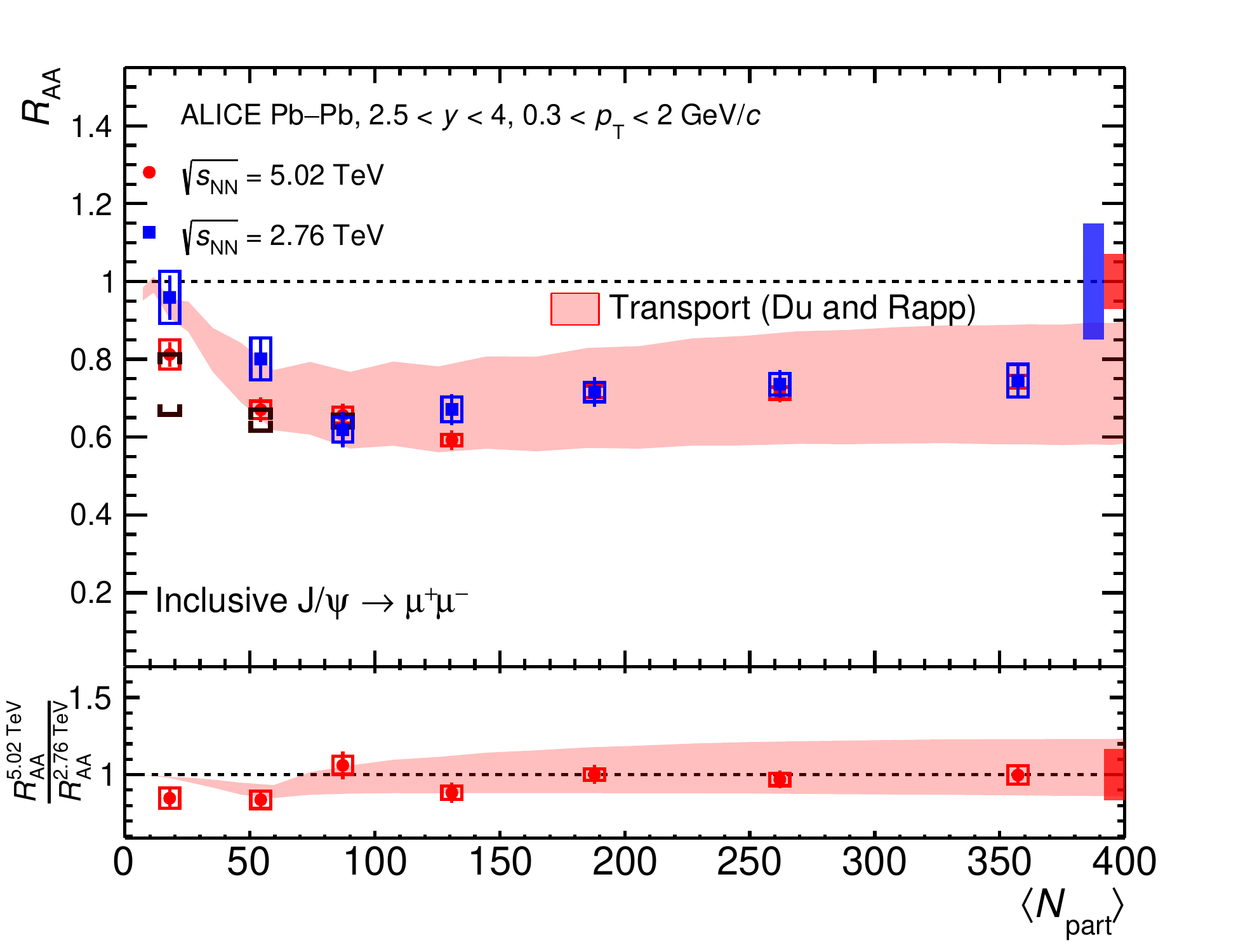}
\includegraphics[width=0.45\textwidth]{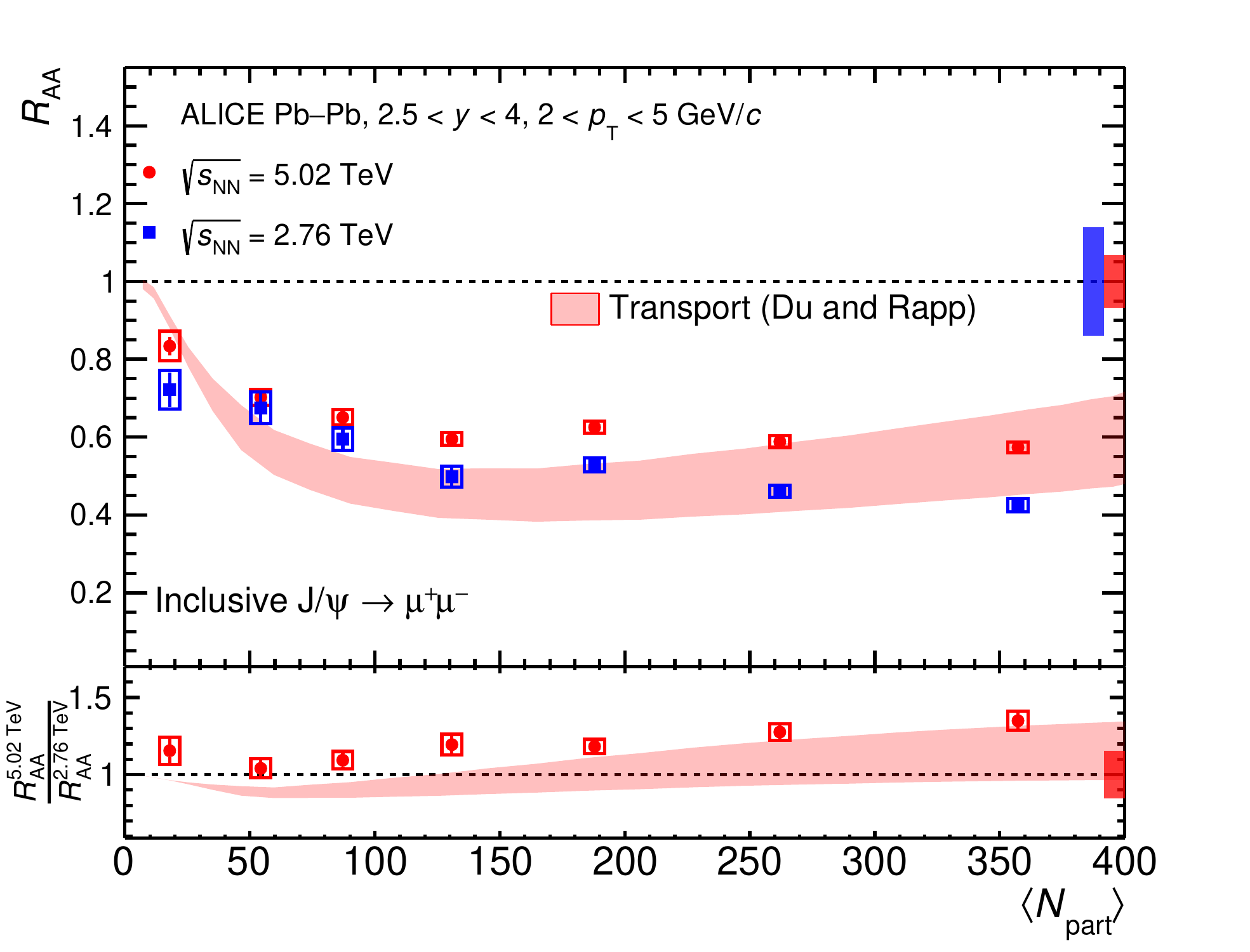}
\includegraphics[width=0.45\textwidth]{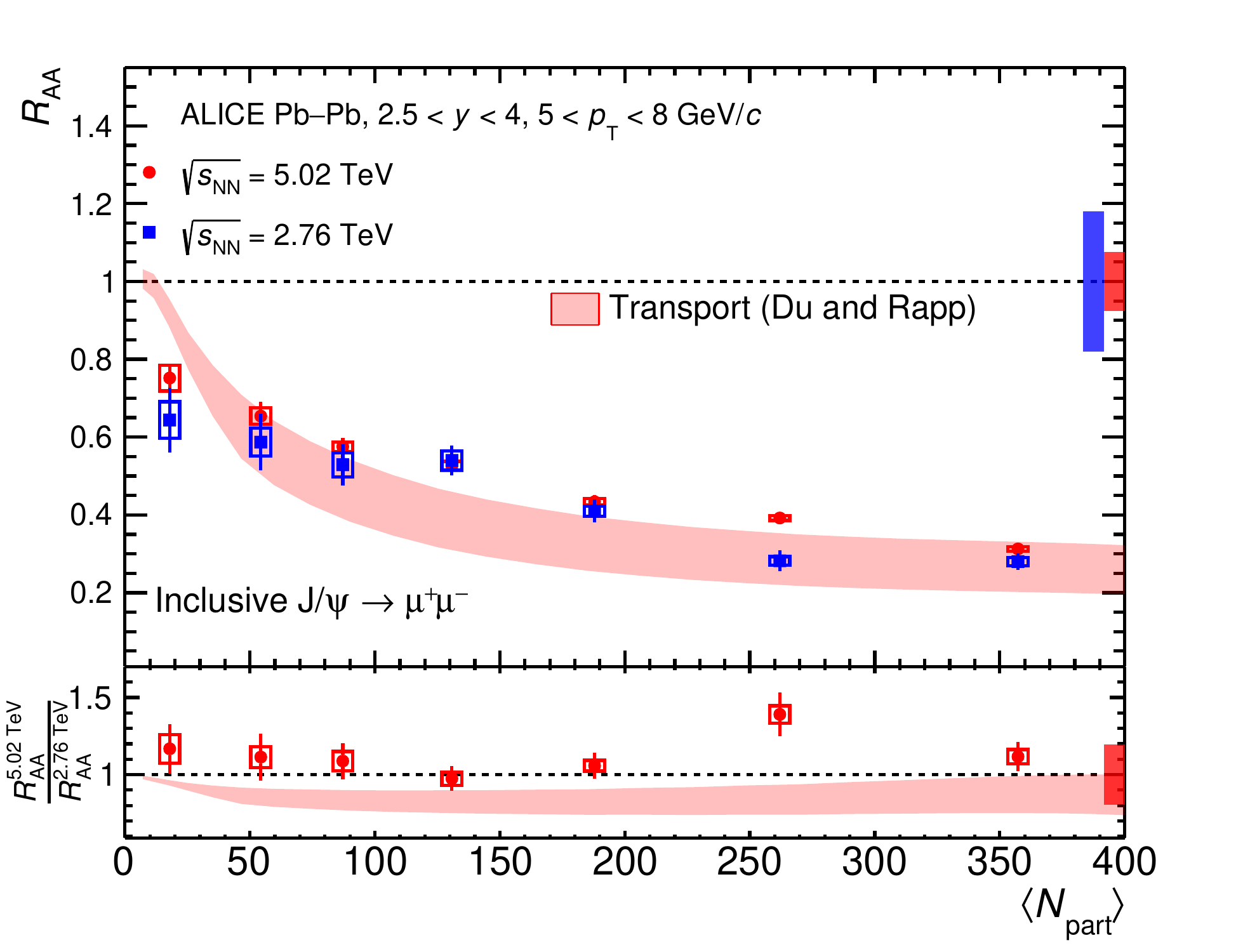}
\includegraphics[width=0.45\textwidth]{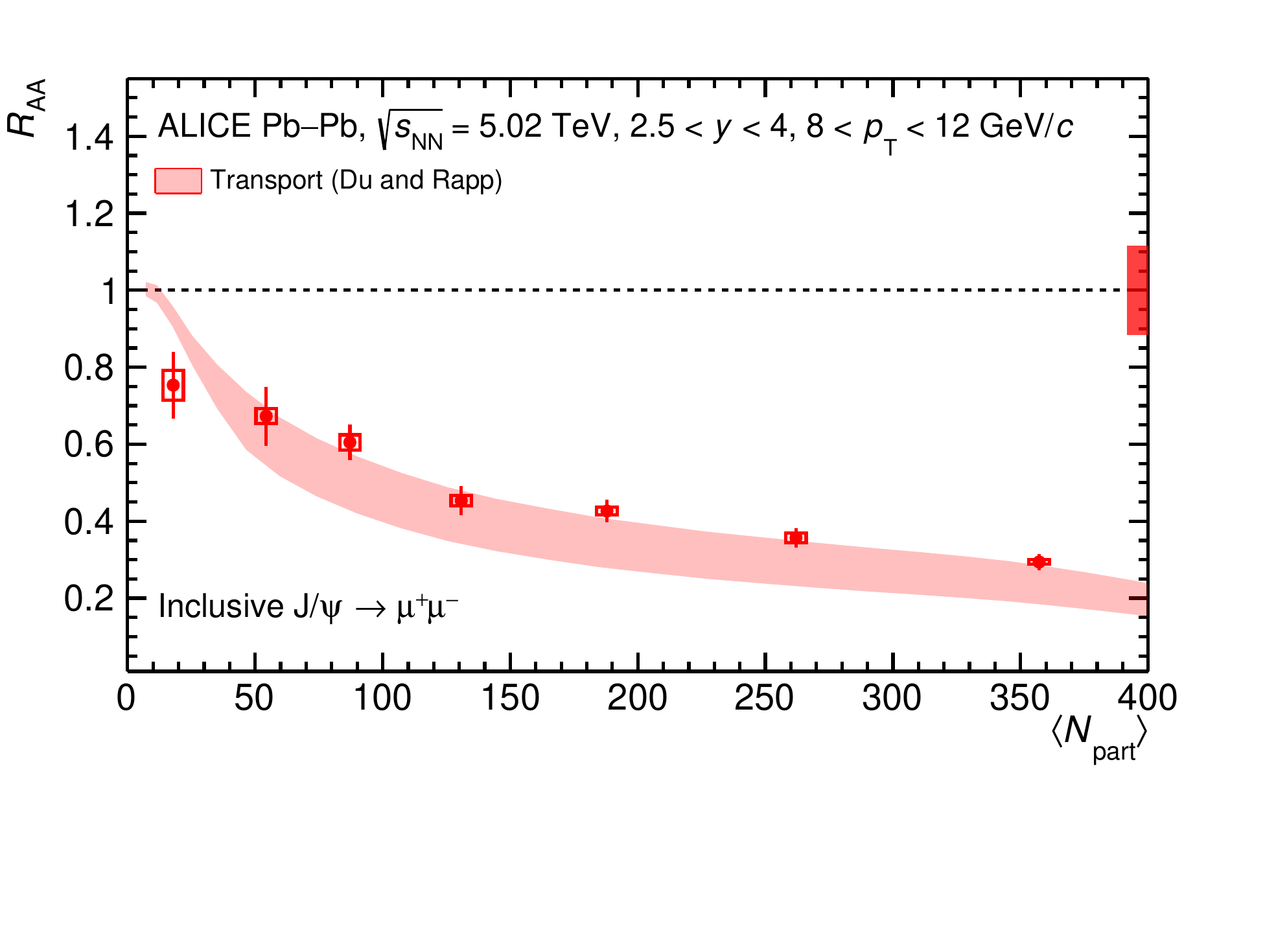}
\caption{Inclusive \jpsi\ nuclear modification factor as a function of \Npart\ for \PbPb\ collisions at \sqrtsnnE{5.02}. Results are shown for four \pt\ intervals. The vertical error bars represent statistical uncertainties, the boxes around the points uncorrelated systematic uncertainties, while the correlated uncertainty is shown as a filled box around $\Raa=1$. When the corresponding results at \sqrtsnnE{2.76} are available, the ratio of the results at the two energies is shown in the bottom section of each figure. The brackets around \Raa\ values for 0.3 $<$ \pt\ $<$ 2 \gevc\ represent an estimate of the influence of \jpsi\ photo-production, as detailed in Sec.~\ref{sec:Raa}. The \Raa\ results at \sqrtsnnE{5.02} as well as the available ratios to the \sqrtsnnE{2.76} results are compared with transport model calculations~\cite{Du:2015wha}.}
\label{fig:6}
\end{center}
\end{figure}

\begin{figure}[!htb]
\begin{center}
\includegraphics[width=0.5\textwidth]{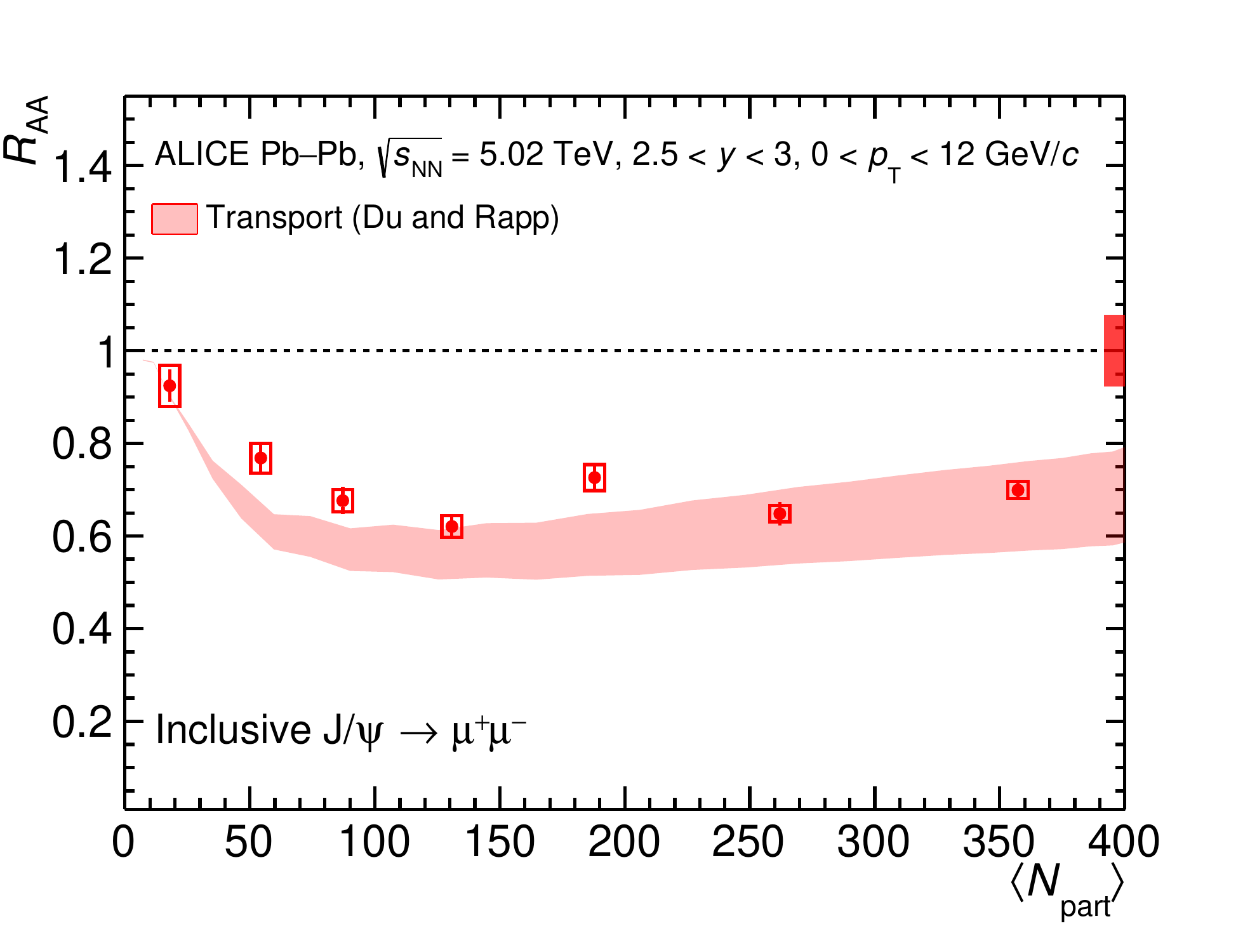}
\includegraphics[width=0.5\textwidth]{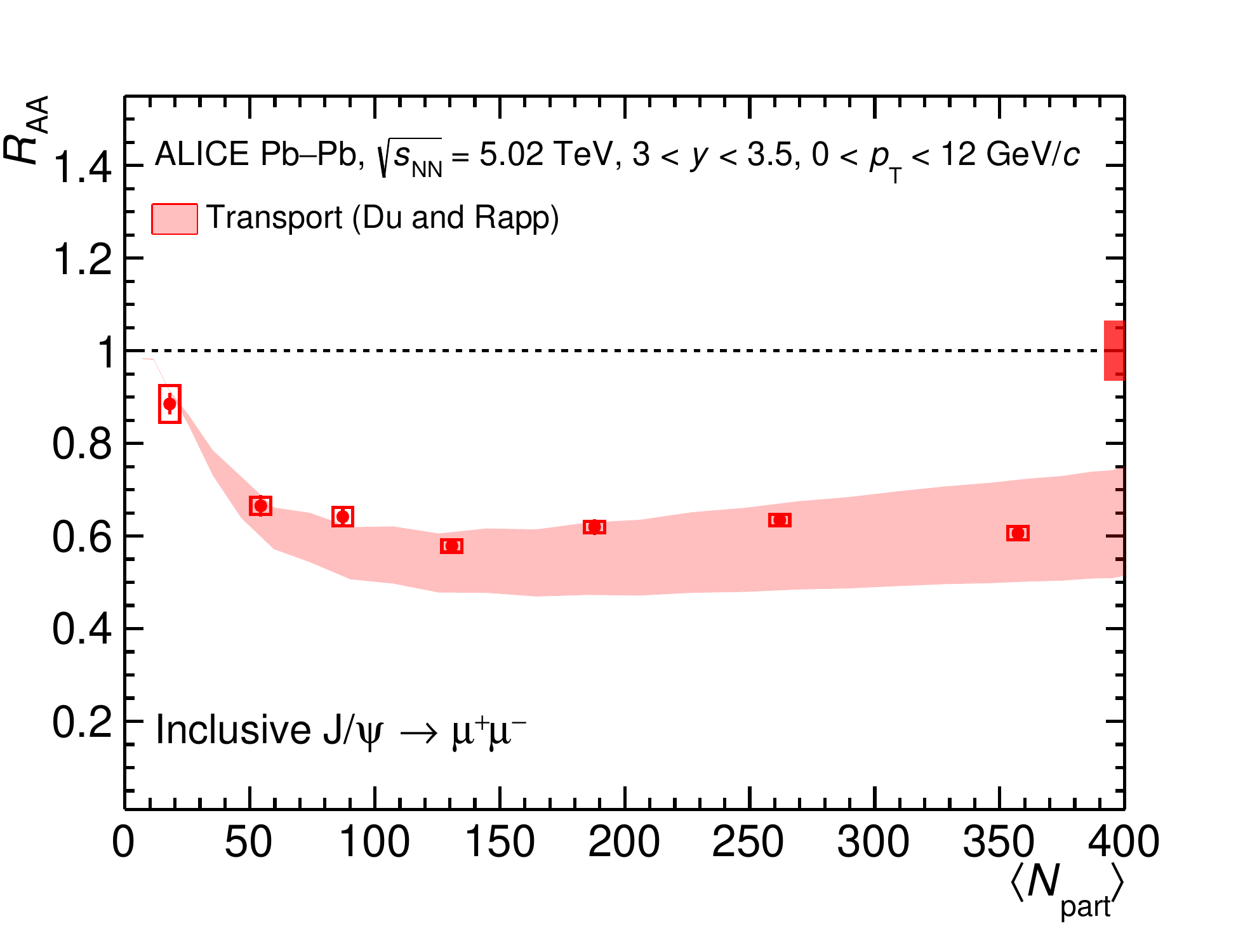}
\includegraphics[width=0.5\textwidth]{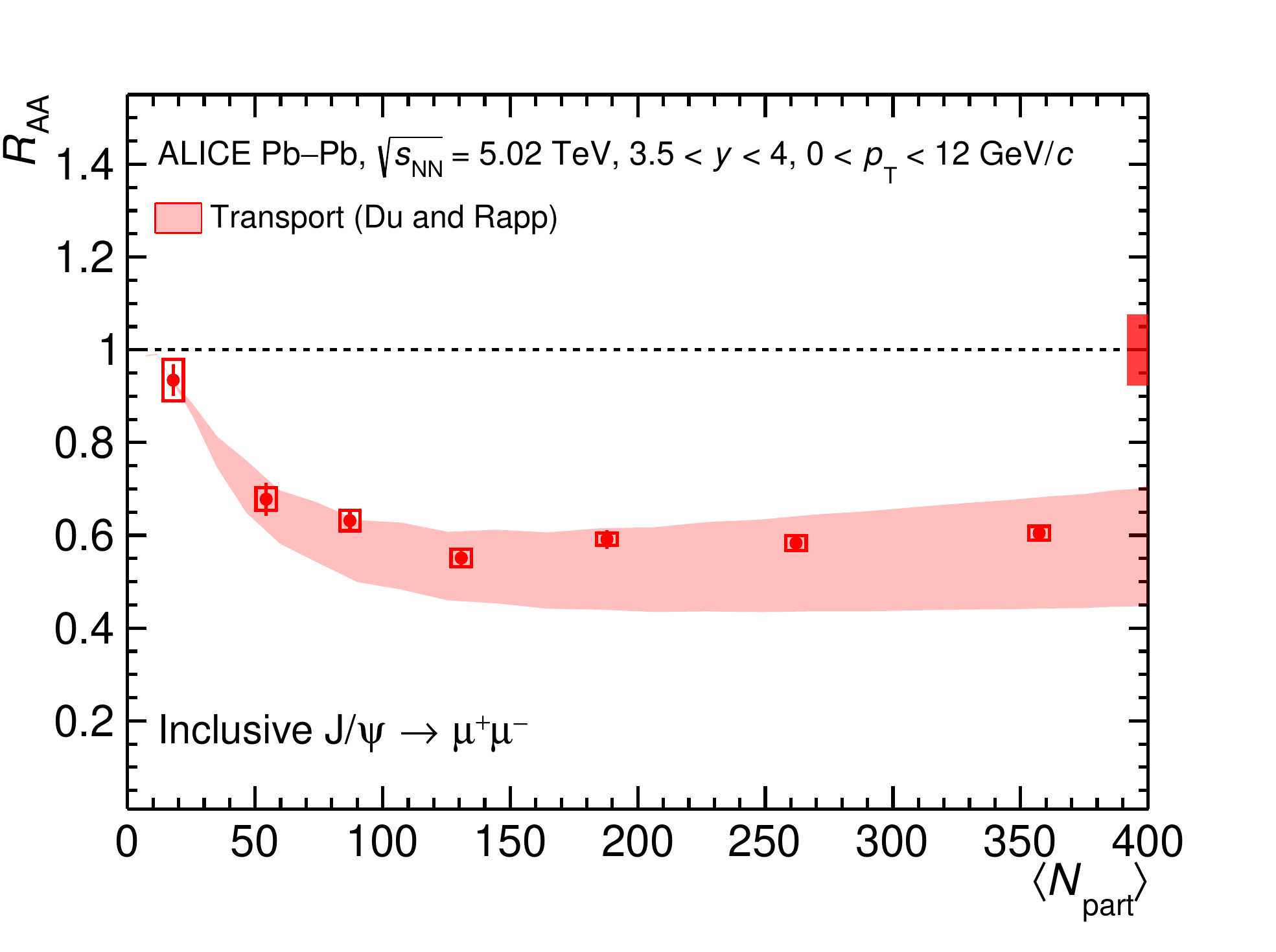}
\caption{Inclusive \jpsi\ nuclear modification factor as a function of \Npart\ for \PbPb\ collisions at \sqrtsnnE{5.02}, in the interval $0.3 < \pt\ < 12$ \gevc. Results are shown for three $y$ intervals. The vertical error bars represent statistical uncertainties, the boxes around the points uncorrelated systematic uncertainties, while the correlated uncertainty is shown as a filled box around $\Raa=1$. The results are compared with transport model calculations~\cite{Du:2015wha}.}
\label{fig:7}
\end{center}
\end{figure}

%	In Fig.~\ref{fig:6} and~\ref{fig:7} the centrality dependence of $R_{\rm AA}$ is shown, for various transverse momentum and  rapidity selections. The centrality intervals correspond to the selections 0--10\%, 10--20\%, 20--30\%, 30--40\%, 40--50\%, 50--60\%, 60--90\%, from central to peripheral events.
In Figs.~\ref{fig:6} and~\ref{fig:7} the \Raa\ as a function of the average number of participant nucleons \Npart\ is shown for various transverse momentum and  rapidity intervals, respectively. The \Npart\ intervals correspond to the centrality selections 0--10\%, 10--20\%, 20--30\%, 30--40\%, 40--50\%, 50--60\%, and 60--90\%, from larger to smaller \Npart\ values.
The results of Fig.~\ref{fig:6} clearly show that moving from low to high \pt\ the centrality dependence of $R_{\rm AA}$ becomes steeper, with $R_{\rm AA}$ reaching a minimum value of $0.29 \pm 0.02 {\rm (stat)} \pm 0.01 {\rm (syst)}$ for the 0--10\% centrality class and 8 $<$ \pt\ $<$ 12 \gevc. In the low-\pt\ region (0.3 $<$ \pt\ $<$ 2 \gevc), the \Raa\ has a weak \Npart\ dependence and is compatible with being constant ($\sim 0.7$) for \Npart\ $>$ 150. In the most peripheral centrality class, a deviation from unity can be observed, in particular for  \pt\ $>$ 2 \gevc, not seen in the theoretical calculations. As discussed in Refs~\cite{Morsch:2017brb,Acharya:2018njl}, the origin may be from the bias introduced by the event selection and collision geometry, which causes an apparent suppression. When comparing the results with those corresponding to \PbPb\ collisions at \sqrtsnnE{2.76}~\cite{Adam:2015isa}, systematically higher $R_{\rm AA}$ values are found in the \pt\ interval 2 $<$ \pt\ $<$ 5 \gevc, even if the maximum observed difference is only at 1.5$\sigma$ level, for the centrality region 0--10\%. In all other \pt\ intervals where the comparison is possible, the results at the two energies are compatible.
When comparing the results with the transport model calculations, the agreement is good at low \pt\ (0.3 $<$ \pt\ $<$ 2 \gevc), while the data lie close to the upper edge of the calculation at higher \pt.

In Fig.~\ref{fig:7} the centrality dependence of the nuclear modification factor is shown for 3 rapidity intervals. No variation of the suppression pattern against rapidity is observed. The same weak dependence can also be observed with the transport model calculations.

\subsection{\jpsi\ average transverse momentum and \rAA}
\label{sec:meanPt}

A complementary insight into the modification of \jpsi\ transverse momentum distributions in \PbPb\ collisions can be obtained by the study of the \jpsi\ average transverse momentum \meanpt\ and the average squared momentum \meanptsq\ as a function of the collision centrality. By normalizing \meanptsq\ to the corresponding pp value, one obtains an adimensional quantity, $r_{\rm AA}=\langle{p_{\rm T}}^2\rangle_{\rm AA}/\langle{p_{\rm T}}^2\rangle_{\rm pp}$, useful for comparisons between various collision energies and/or theory calculations.

\begin{figure}[!htb]
			\centering
			\includegraphics[width=0.7\textwidth]{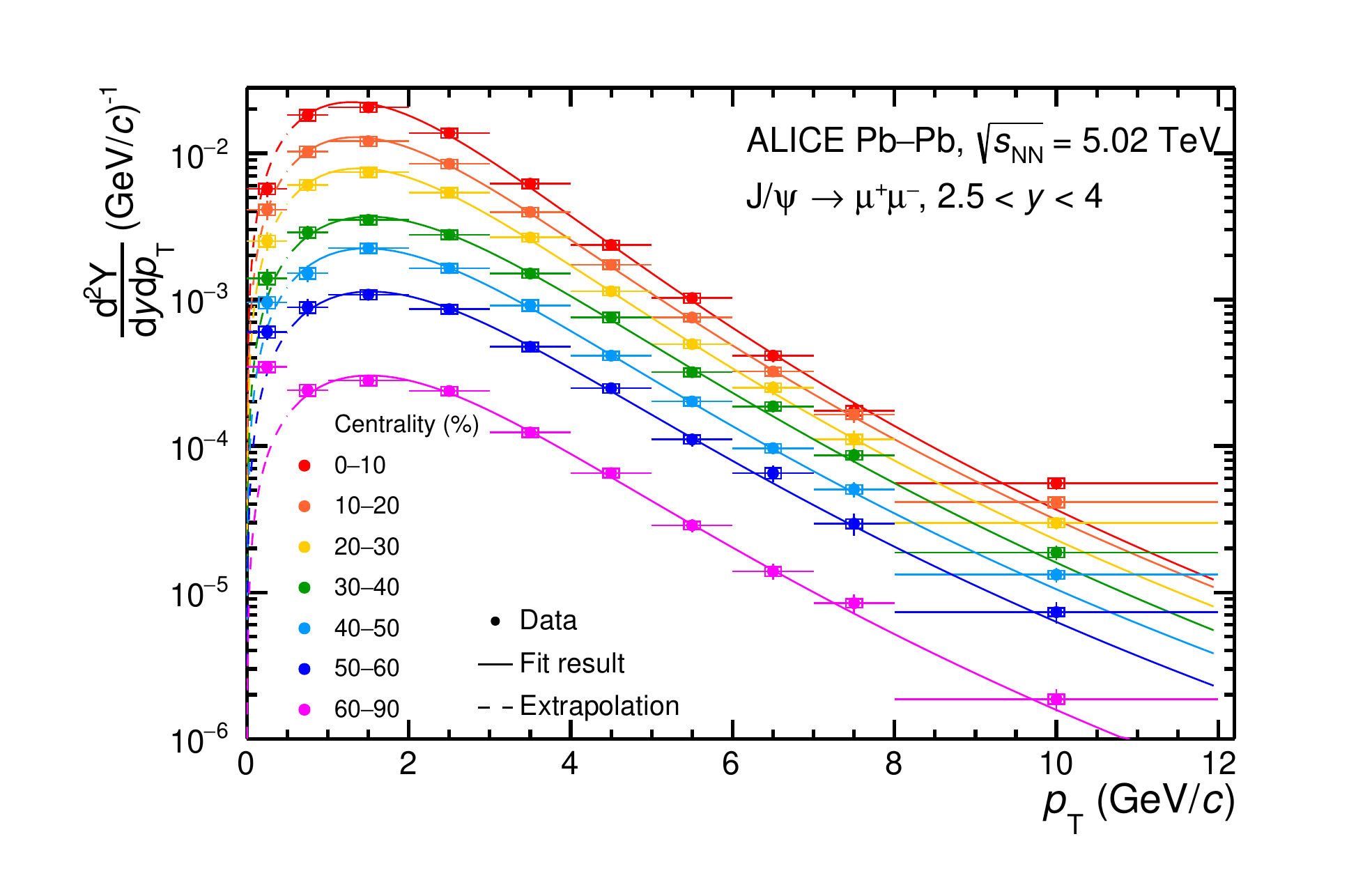}
			\caption{Inclusive \jpsi\ yields as a function of \pt\ in \PbPb\ collisions at \sqrtsnnE{5.02}, for various centrality classes. The vertical error bars represent the statistical uncertainties while the uncorrelated systematic uncertainties are represented by boxes around the points. The curves are the results of fits obtained using the function shown in Eq.~\ref{eq:fPt}. The dashed region corresponds to the region \pt\ $<$ 0.5 \gevc, excluded in the fits. }
\label{fig:8}
\end{figure}

As a first step, the \jpsi\ invariant yields as a function of \pt\ are fitted in various centrality classes with the following function

\begin{equation}\label{eq:fPt}
f(\pt) = C \cdot \frac{\pt}{\Big( 1+(\pt/p_0)^2 \Big)^n},
\end{equation}

where $C$, $p_0$ and $n$ are free parameters. This function is widely used to reproduce the \jpsi\ \pt\ distribution in hadronic collisions (e.g Refs.~\cite{PhysRevLett.41.684,Bossu:2011qe}). The quantities to be determined, \meanpt\ and \meanptsq, are then computed as the first and second moment of $f(\pt)$ respectively. In Fig.~\ref{fig:8}, the \jpsi\ invariant yields as a function of \pt\ are shown for various centrality classes together with the fitted functions. In order to limit the influence of the \jpsi\ production excess at low \pt, due to photo-production, the interval \pt\ $<$ 0.5 \gevc\ was excluded from the fit. The statistical (systematic) uncertainties on \meanpt\ and \meanptsq\ were obtained from fits to the invariant yield distributions, considering only statistical (\pt-uncorrelated systematic) uncertainties on the \jpsi\ yields.

\begin{figure}[!htb]
\centering		
\includegraphics[width=0.49\textwidth]{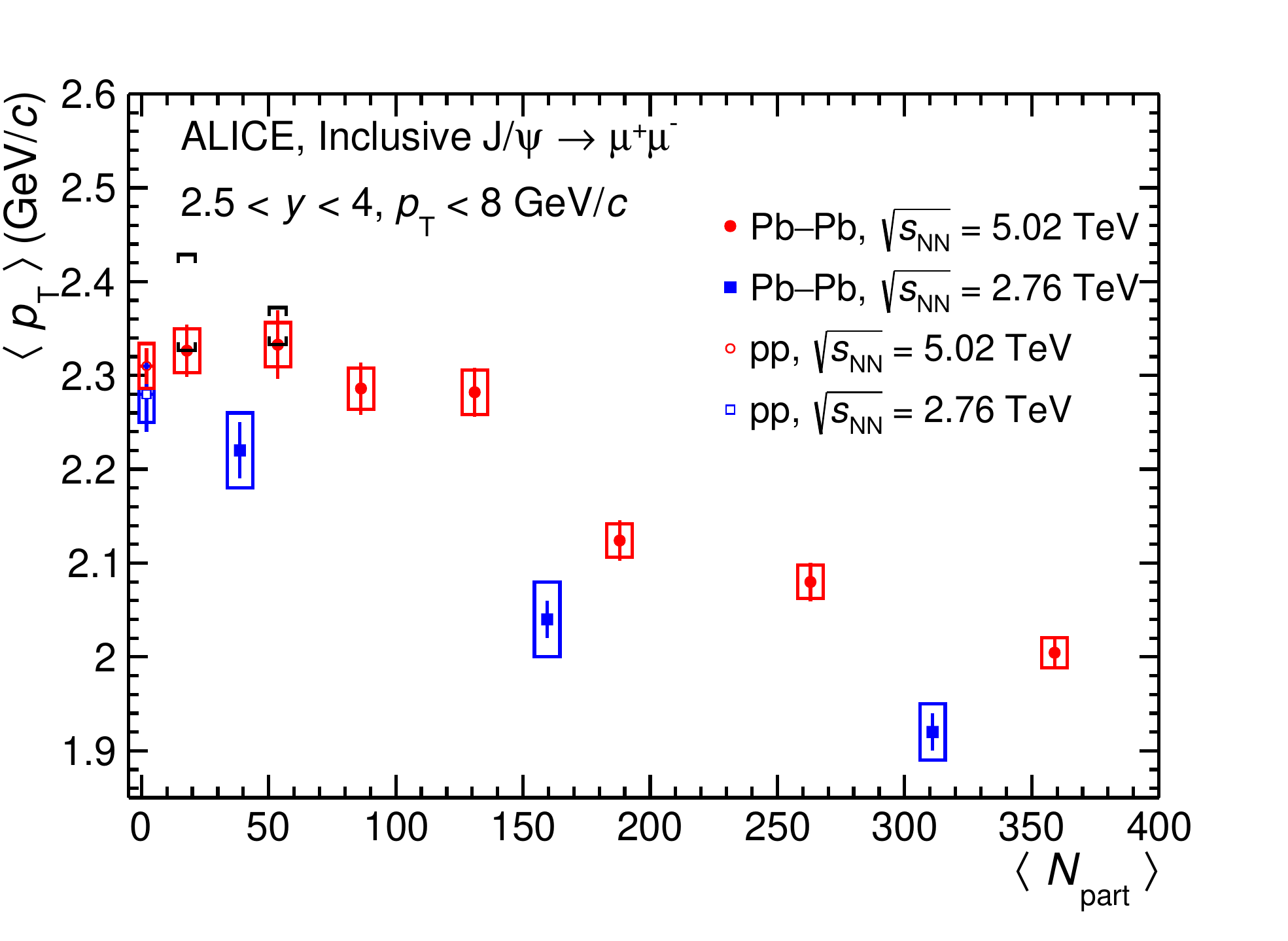}
\includegraphics[width=0.49\textwidth]{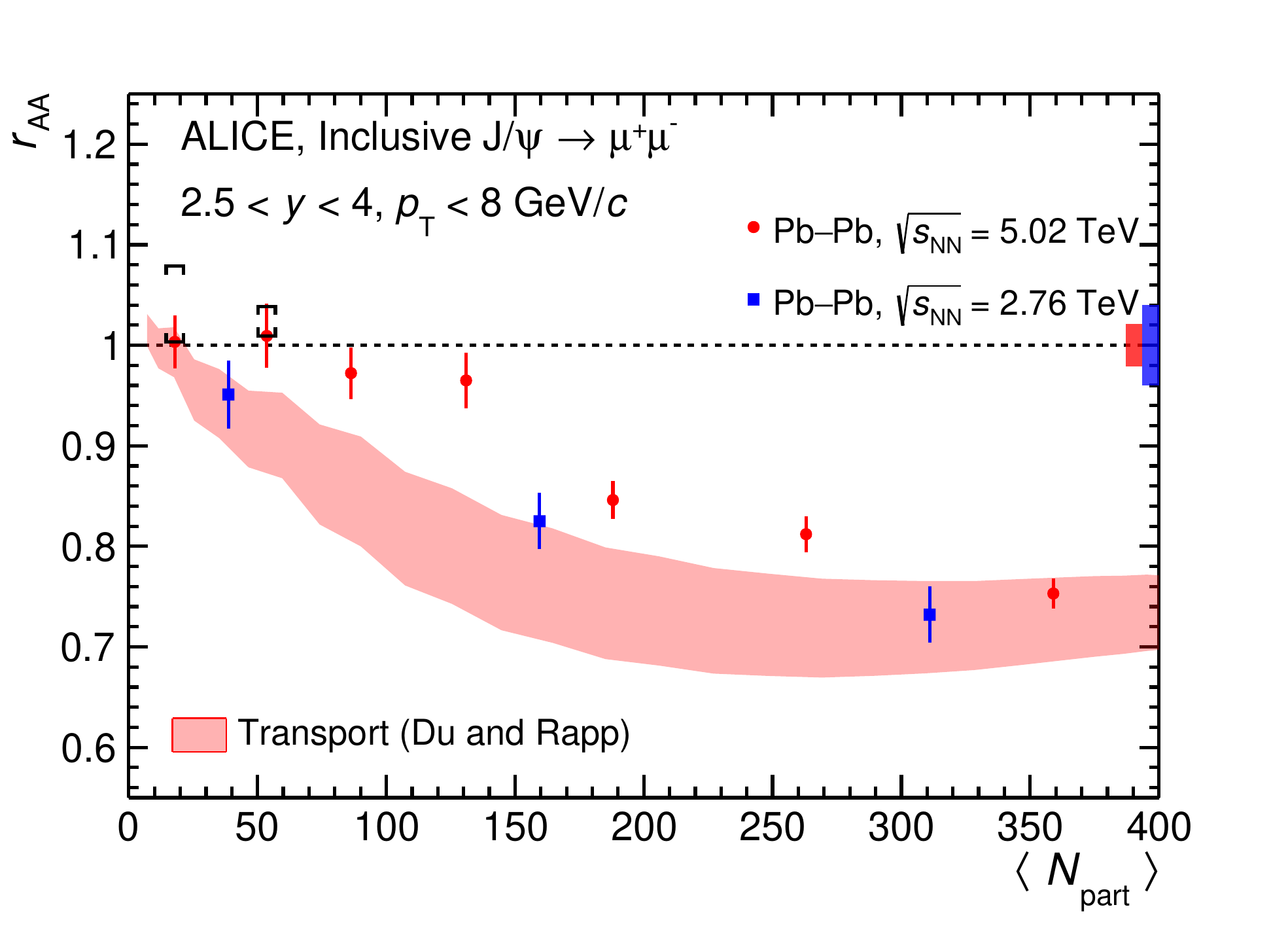}
\caption{(Left) Inclusive \jpsi\ \meanpt\ measured in \PbPb\ collisions at \sqrtsnnE{5.02} and \sqrtsnnE{2.76}, as a function of \Npart\ for \pt\ $<$ 8 \gevc. The vertical error bars represent the statistical uncertainties, while the uncorrelated systematic uncertainties are shown as boxes around the points. \jpsi\ \meanpt\ results in \pp\ collisions at the two collision energies are also shown at \Npart = 2. (Right) Inclusive \jpsi\ \rAA\ in \PbPb\ collisions at \sqrtsnnE{5.02}, compared with the \sqrtsnnE{2.76} results and a transport model calculation~\cite{Du:2015wha}, as a function of \Npart\ for \pt\ $<$ 8 \gevc. The 
vertical error bars represent the quadratic sum of statistical and systematic uncertainties on the numerator of the $r_{\rm AA}$ expression ($\langle{p_{\rm T}}^2\rangle_{\rm AA}$), while the uncertainties on the denominator ($\langle{p_{\rm T}}^2\rangle_{\rm pp}$) are shown as a filled box around unity. In the two panels,  the brackets around the two most peripheral data points represent an estimate of the maximum influence of \jpsi\ photo-production, as detailed in Sec.~\ref{sec:meanPt}.}
\label{fig:9}
\end{figure}

In the left panel of Fig.~\ref{fig:9}, the centrality dependence of \meanpt\ is shown and compared with previous results at \sqrtsnnE{2.76}~\cite{Adam:2015isa}. The centrality dependence of the \sqrtsnnE{5.02} results is weak up to \Npart\ $\sim$ 150, followed by a significant decrease towards central events. This softening of the \jpsi\ \pt\ distributions is a direct consequence of the smaller suppression observed at low \pt\ when considering the transverse-momentum dependence of the nuclear modification factors, shown in Fig.~\ref{fig:4}. The \meanpt\ values are systematically larger than those at \sqrtsnnE{2.76}, an effect due to the increase of the collision energy, but the decrease of \meanpt\ with increasing centrality is similar at the two energies. A more direct comparison with lower energy results and theoretical calculations can be performed by studying the quantity \rAA. The results are shown in the right panel of Fig.~\ref{fig:9}, and compared with those obtained in \PbPb\ collisions at \sqrtsnnE{2.76} and the transport model calculations. In peripheral collisions, and up to \Npart\ $\sim$ 150, the \rAA\ value is compatible with unity within uncertainties. A maximum decrease of $\sim$25\% is observed for central collisions. The brackets around the \meanpt\ and \rAA\ in peripheral collisions represent the possible variation of the hadronic \jpsi\ \meanpt\ and \rAA\ for two extreme hypotheses on the \jpsi\ photo-production contamination. The lower limit bracket corresponds to the assumption of no contribution from photo-produced \jpsi, while the upper one corresponds to the hypothesis that all the \jpsi\ with \pt\ $<$ 300 \mevc\ are photo-produced.
The results are compatible with those at \sqrtsnnE{2.76}, with a hint for larger \rAA\ values at \sqrtsnnE{5.02}. A very different centrality dependence was observed at lower collision energies ($\sqrtsnn$ $=$ 200 GeV at RHIC~\cite{PhysRevLett.101.122301} and $\sqrtsnn$ $=$ 17 GeV at SPS~\cite{ABREU200185}) as the \rAA\ increases (especially at the SPS energy) towards more central collisions (the comparison was shown in Ref.~\cite{Adam:2015isa}). The different behaviors of \rAA\ at different energies can be explained by the increasing amount of \jpsi\ regeneration with collision energy. Finally, the comparison with the transport model calculation ~\cite{Du:2015wha} shows good agreement for peripheral and central collisions, but an underestimation of the data points is observed in the intermediate centrality class, reaching a significance up to 2.5$\sigma$ for \Npart~$\sim$~150.  

\section{Conclusions}\label{sec:conclusions}

This paper reports on ALICE measurements of the inclusive \jpsi\ production in \PbPb\ collisions at \sqrtsnnE{5.02} in the kinematic range 2.5 $<$ \y\ $<$ 4 up to \pt\ $<$ 12 \gevc. Results on the nuclear modification factor \Raa, the average transverse-momentum \meanpt, and the ratio \rAA\ were presented. A systematic comparison with the calculation of a transport model was carried out and, for the \pt\ dependence of \Raa, with the results of a statistical hadronization model.

The inclusive \jpsi\ \Raa\ as a function of transverse momentum and rapidity for the centrality range 0--90\%, is compatible with previously published results at \sqrtsnnE{2.76}~\cite{Adam:2015isa}. A suppression of the \jpsi\ production is observed (\Raa$< 1$), mild at low \pt\ but increasing towards higher \pt, and not strongly depending on rapidity. The centrality-differential studies show that the \y\ dependence of \Raa\ is weak and fairly independent of centrality, while the \pt\ dependence of \Raa\ grows steeper for more central events. All the \Raa\ results are fairly reproduced by the calculation of a transport model, with a tendency to underestimate the observed \Raa\ at intermediate \pt. The statistical hadronization model reproduces, although with larger uncertainties, the \pt\ dependence of \Raa\ for various centrality classes, but shows a discrepancy in the high-\pt\ region.

%When studying the \pt-dependence of the \Raa\ for different centrality classes, the increasing trend of $R_{\rm AA}$ at low \pt\ smoothly disappears when moving from central to peripheral events. As a function of $y$, the \Raa\ remains fairly constant from central to peripheral events.

A complementary study was also carried out by measuring the centrality dependence of $R_{\rm AA}$ for different \pt\ and $y$ intervals. A suppression strongly increasing with centrality is visible at high \pt, while at low \pt\ the suppression is relatively weak ($R_{\rm AA}\sim 0.7$) and practically independent of centrality. On the contrary, the shape of the \Raa\ as a function of centrality does not vary significantly in the studied rapidity ranges, showing a mild decrease until $\Npart \sim 100$, followed by a plateau.

Finally, the $r_{\rm AA}$ ratio decreases with increasing centrality, similarly to previous observations at \sqrtsnnE{2.76}. The transport model calculation underestimates the measurement at intermediate \Npart\ values.
%%A tension with the results of the transport model can be observed at intermediate $N_{\rm part}$ values.

The results shown in this paper confirm, with better accuracy, the observations carried out at \sqrtsnnE{2.76} and strengthen the evidence for the presence of a mechanism that leads to a significant increase of $R_{\rm AA}$ at low \pt. Recombination of charm-quark pairs during the deconfined QGP phase, as implemented in the transport model compared with our results, is a strong candidate for explaining the features of the data.

\newenvironment{acknowledgement}{\relax}{\relax}
\begin{acknowledgement}
\section*{Acknowledgements}
% Version: 2019-09-05

The ALICE Collaboration would like to thank all its engineers and technicians for their invaluable contributions to the construction of the experiment and the CERN accelerator teams for the outstanding performance of the LHC complex.
The ALICE Collaboration gratefully acknowledges the resources and support provided by all Grid centres and the Worldwide LHC Computing Grid (WLCG) collaboration.
The ALICE Collaboration acknowledges the following funding agencies for their support in building and running the ALICE detector:
A. I. Alikhanyan National Science Laboratory (Yerevan Physics Institute) Foundation (ANSL), State Committee of Science and World Federation of Scientists (WFS), Armenia;
Austrian Academy of Sciences, Austrian Science Fund (FWF): [M 2467-N36] and Nationalstiftung f\"{u}r Forschung, Technologie und Entwicklung, Austria;
Ministry of Communications and High Technologies, National Nuclear Research Center, Azerbaijan;
Conselho Nacional de Desenvolvimento Cient\'{\i}fico e Tecnol\'{o}gico (CNPq), Financiadora de Estudos e Projetos (Finep), Funda\c{c}\~{a}o de Amparo \`{a} Pesquisa do Estado de S\~{a}o Paulo (FAPESP) and Universidade Federal do Rio Grande do Sul (UFRGS), Brazil;
Ministry of Education of China (MOEC) , Ministry of Science \& Technology of China (MSTC) and National Natural Science Foundation of China (NSFC), China;
Ministry of Science and Education and Croatian Science Foundation, Croatia;
Centro de Aplicaciones Tecnol\'{o}gicas y Desarrollo Nuclear (CEADEN), Cubaenerg\'{\i}a, Cuba;
Ministry of Education, Youth and Sports of the Czech Republic, Czech Republic;
The Danish Council for Independent Research | Natural Sciences, the VILLUM FONDEN and Danish National Research Foundation (DNRF), Denmark;
Helsinki Institute of Physics (HIP), Finland;
Commissariat \`{a} l'Energie Atomique (CEA), Institut National de Physique Nucl\'{e}aire et de Physique des Particules (IN2P3) and Centre National de la Recherche Scientifique (CNRS) and R\'{e}gion des  Pays de la Loire, France;
Bundesministerium f\"{u}r Bildung und Forschung (BMBF) and GSI Helmholtzzentrum f\"{u}r Schwerionenforschung GmbH, Germany;
General Secretariat for Research and Technology, Ministry of Education, Research and Religions, Greece;
National Research, Development and Innovation Office, Hungary;
Department of Atomic Energy Government of India (DAE), Department of Science and Technology, Government of India (DST), University Grants Commission, Government of India (UGC) and Council of Scientific and Industrial Research (CSIR), India;
Indonesian Institute of Science, Indonesia;
Centro Fermi - Museo Storico della Fisica e Centro Studi e Ricerche Enrico Fermi and Istituto Nazionale di Fisica Nucleare (INFN), Italy;
Institute for Innovative Science and Technology , Nagasaki Institute of Applied Science (IIST), Japanese Ministry of Education, Culture, Sports, Science and Technology (MEXT) and Japan Society for the Promotion of Science (JSPS) KAKENHI, Japan;
Consejo Nacional de Ciencia (CONACYT) y Tecnolog\'{i}a, through Fondo de Cooperaci\'{o}n Internacional en Ciencia y Tecnolog\'{i}a (FONCICYT) and Direcci\'{o}n General de Asuntos del Personal Academico (DGAPA), Mexico;
Nederlandse Organisatie voor Wetenschappelijk Onderzoek (NWO), Netherlands;
The Research Council of Norway, Norway;
Commission on Science and Technology for Sustainable Development in the South (COMSATS), Pakistan;
Pontificia Universidad Cat\'{o}lica del Per\'{u}, Peru;
Ministry of Science and Higher Education and National Science Centre, Poland;
Korea Institute of Science and Technology Information and National Research Foundation of Korea (NRF), Republic of Korea;
Ministry of Education and Scientific Research, Institute of Atomic Physics and Ministry of Research and Innovation and Institute of Atomic Physics, Romania;
Joint Institute for Nuclear Research (JINR), Ministry of Education and Science of the Russian Federation, National Research Centre Kurchatov Institute, Russian Science Foundation and Russian Foundation for Basic Research, Russia;
Ministry of Education, Science, Research and Sport of the Slovak Republic, Slovakia;
National Research Foundation of South Africa, South Africa;
Swedish Research Council (VR) and Knut \& Alice Wallenberg Foundation (KAW), Sweden;
European Organization for Nuclear Research, Switzerland;
Suranaree University of Technology (SUT), National Science and Technology Development Agency (NSDTA) and Office of the Higher Education Commission under NRU project of Thailand, Thailand;
Turkish Atomic Energy Agency (TAEK), Turkey;
National Academy of  Sciences of Ukraine, Ukraine;
Science and Technology Facilities Council (STFC), United Kingdom;
National Science Foundation of the United States of America (NSF) and United States Department of Energy, Office of Nuclear Physics (DOE NP), United States of America.    %%%%%%% done by webmaster team
\end{acknowledgement}

%%%%%%%% Bibliography (In case of using bibtex generate the bbl requested by arXiv)
\bibliographystyle{utphys}   % Remember we use title in the biblio
\bibliography{bibliography}
% \input{bibliography.tex}

%%%%%%%%% appendix with author list
\clearpage
\appendix
\section{The ALICE Collaboration}
\label{app:collab}
% Collaboration: CERN-LHC-ALICE
% Generation Date is 2019-Jul-08

% How to use:
%%%%%%%%% appendix with author list
%\appendix
%\section{The ALICE Collaboration}
%\label{app:collab}
%\input{Alice_Authorslist_XXXX-Axx-XX.tex}
\begingroup
\small
\begin{flushleft}
S.~Acharya\Irefn{org141}\And 
D.~Adamov\'{a}\Irefn{org93}\And 
S.P.~Adhya\Irefn{org141}\And 
A.~Adler\Irefn{org73}\And 
J.~Adolfsson\Irefn{org79}\And 
M.M.~Aggarwal\Irefn{org98}\And 
G.~Aglieri Rinella\Irefn{org34}\And 
M.~Agnello\Irefn{org31}\And 
N.~Agrawal\Irefn{org10}\textsuperscript{,}\Irefn{org48}\textsuperscript{,}\Irefn{org53}\And 
Z.~Ahammed\Irefn{org141}
% ...
\And
\renewcommand\labelenumi{\textsuperscript{\theenumi}~}

\section*{Affiliation notes}
\renewcommand\theenumi{\roman{enumi}}
\begin{Authlist}
\item \Adef{org*}Deceased
\item \Adef{orgI}Dipartimento DET del Politecnico di Torino, Turin, Italy
\item \Adef{orgII}M.V. Lomonosov Moscow State University, D.V. Skobeltsyn Institute of Nuclear, Physics, Moscow, Russia
\item \Adef{orgIII}Department of Applied Physics, Aligarh Muslim University, Aligarh, India
\item \Adef{orgIV}Institute of Theoretical Physics, University of Wroclaw, Poland
\end{Authlist}

\section*{Collaboration Institutes}
\renewcommand\theenumi{\arabic{enumi}~}
\begin{Authlist}
\item \Idef{org1}A.I. Alikhanyan National Science Laboratory (Yerevan Physics Institute) Foundation, Yerevan, Armenia
\item \Idef{org2}Bogolyubov Institute for Theoretical Physics, National Academy of Sciences of Ukraine, Kiev, Ukraine
\item \Idef{org3}Bose Institute, Department of Physics  and Centre for Astroparticle Physics and Space Science (CAPSS), Kolkata, India
\item \Idef{org4}Budker Institute for Nuclear Physics, Novosibirsk, Russia
\item \Idef{org5}California Polytechnic State University, San Luis Obispo, California, United States
\item \Idef{org6}Central China Normal University, Wuhan, China
\item \Idef{org7}Centre de Calcul de l'IN2P3, Villeurbanne, Lyon, France
\item \Idef{org8}Centro de Aplicaciones Tecnol\'{o}gicas y Desarrollo Nuclear (CEADEN), Havana, Cuba
\item \Idef{org9}Centro de Investigaci\'{o}n y de Estudios Avanzados (CINVESTAV), Mexico City and M\'{e}rida, Mexico
\item \Idef{org10}Centro Fermi - Museo Storico della Fisica e Centro Studi e Ricerche ``Enrico Fermi', Rome, Italy
\item \Idef{org11}Chicago State University, Chicago, Illinois, United States
\item \Idef{org12}China Institute of Atomic Energy, Beijing, China
\item \Idef{org13}Chonbuk National University, Jeonju, Republic of Korea
\item \Idef{org14}Comenius University Bratislava, Faculty of Mathematics, Physics and Informatics, Bratislava, Slovakia
\item \Idef{org15}COMSATS University Islamabad, Islamabad, Pakistan
\item \Idef{org16}Creighton University, Omaha, Nebraska, United States
\item \Idef{org17}Department of Physics, Aligarh Muslim University, Aligarh, India
\item \Idef{org18}Department of Physics, Pusan National University, Pusan, Republic of Korea
\item \Idef{org19}Department of Physics, Sejong University, Seoul, Republic of Korea
\item \Idef{org20}Department of Physics, University of California, Berkeley, California, United States
\item \Idef{org21}Department of Physics, University of Oslo, Oslo, Norway
\item \Idef{org22}Department of Physics and Technology, University of Bergen, Bergen, Norway
\item \Idef{org23}Dipartimento di Fisica dell'Universit\`{a} 'La Sapienza' and Sezione INFN, Rome, Italy
\item \Idef{org24}Dipartimento di Fisica dell'Universit\`{a} and Sezione INFN, Cagliari, Italy
\item \Idef{org25}Dipartimento di Fisica dell'Universit\`{a} and Sezione INFN, Trieste, Italy
\item \Idef{org26}Dipartimento di Fisica dell'Universit\`{a} and Sezione INFN, Turin, Italy
\item \Idef{org27}Dipartimento di Fisica e Astronomia dell'Universit\`{a} and Sezione INFN, Bologna, Italy
\item \Idef{org28}Dipartimento di Fisica e Astronomia dell'Universit\`{a} and Sezione INFN, Catania, Italy
\item \Idef{org29}Dipartimento di Fisica e Astronomia dell'Universit\`{a} and Sezione INFN, Padova, Italy
\item \Idef{org30}Dipartimento di Fisica `E.R.~Caianiello' dell'Universit\`{a} and Gruppo Collegato INFN, Salerno, Italy
\item \Idef{org31}Dipartimento DISAT del Politecnico and Sezione INFN, Turin, Italy
\item \Idef{org32}Dipartimento di Scienze e Innovazione Tecnologica dell'Universit\`{a} del Piemonte Orientale and INFN Sezione di Torino, Alessandria, Italy
\item \Idef{org33}Dipartimento Interateneo di Fisica `M.~Merlin' and Sezione INFN, Bari, Italy
\item \Idef{org34}European Organization for Nuclear Research (CERN), Geneva, Switzerland
\item \Idef{org35}Faculty of Electrical Engineering, Mechanical Engineering and Naval Architecture, University of Split, Split, Croatia
\item \Idef{org36}Faculty of Engineering and Science, Western Norway University of Applied Sciences, Bergen, Norway
\item \Idef{org37}Faculty of Nuclear Sciences and Physical Engineering, Czech Technical University in Prague, Prague, Czech Republic
\item \Idef{org38}Faculty of Science, P.J.~\v{S}af\'{a}rik University, Ko\v{s}ice, Slovakia
\item \Idef{org39}Frankfurt Institute for Advanced Studies, Johann Wolfgang Goethe-Universit\"{a}t Frankfurt, Frankfurt, Germany
\item \Idef{org40}Gangneung-Wonju National University, Gangneung, Republic of Korea
\item \Idef{org41}Gauhati University, Department of Physics, Guwahati, India
\item \Idef{org42}Helmholtz-Institut f\"{u}r Strahlen- und Kernphysik, Rheinische Friedrich-Wilhelms-Universit\"{a}t Bonn, Bonn, Germany
\item \Idef{org43}Helsinki Institute of Physics (HIP), Helsinki, Finland
\item \Idef{org44}High Energy Physics Group,  Universidad Aut\'{o}noma de Puebla, Puebla, Mexico
\item \Idef{org45}Hiroshima University, Hiroshima, Japan
\item \Idef{org46}Hochschule Worms, Zentrum  f\"{u}r Technologietransfer und Telekommunikation (ZTT), Worms, Germany
\item \Idef{org47}Horia Hulubei National Institute of Physics and Nuclear Engineering, Bucharest, Romania
\item \Idef{org48}Indian Institute of Technology Bombay (IIT), Mumbai, India
\item \Idef{org49}Indian Institute of Technology Indore, Indore, India
\item \Idef{org50}Indonesian Institute of Sciences, Jakarta, Indonesia
\item \Idef{org51}INFN, Laboratori Nazionali di Frascati, Frascati, Italy
\item \Idef{org52}INFN, Sezione di Bari, Bari, Italy
\item \Idef{org53}INFN, Sezione di Bologna, Bologna, Italy
\item \Idef{org54}INFN, Sezione di Cagliari, Cagliari, Italy
\item \Idef{org55}INFN, Sezione di Catania, Catania, Italy
\item \Idef{org56}INFN, Sezione di Padova, Padova, Italy
\item \Idef{org57}INFN, Sezione di Roma, Rome, Italy
\item \Idef{org58}INFN, Sezione di Torino, Turin, Italy
\item \Idef{org59}INFN, Sezione di Trieste, Trieste, Italy
\item \Idef{org60}Inha University, Incheon, Republic of Korea
\item \Idef{org61}Institut de Physique Nucl\'{e}aire d'Orsay (IPNO), Institut National de Physique Nucl\'{e}aire et de Physique des Particules (IN2P3/CNRS), Universit\'{e} de Paris-Sud, Universit\'{e} Paris-Saclay, Orsay, France
\item \Idef{org62}Institute for Nuclear Research, Academy of Sciences, Moscow, Russia
\item \Idef{org63}Institute for Subatomic Physics, Utrecht University/Nikhef, Utrecht, Netherlands
\item \Idef{org64}Institute of Experimental Physics, Slovak Academy of Sciences, Ko\v{s}ice, Slovakia
\item \Idef{org65}Institute of Physics, Homi Bhabha National Institute, Bhubaneswar, India
\item \Idef{org66}Institute of Physics of the Czech Academy of Sciences, Prague, Czech Republic
\item \Idef{org67}Institute of Space Science (ISS), Bucharest, Romania
\item \Idef{org68}Institut f\"{u}r Kernphysik, Johann Wolfgang Goethe-Universit\"{a}t Frankfurt, Frankfurt, Germany
\item \Idef{org69}Instituto de Ciencias Nucleares, Universidad Nacional Aut\'{o}noma de M\'{e}xico, Mexico City, Mexico
\item \Idef{org70}Instituto de F\'{i}sica, Universidade Federal do Rio Grande do Sul (UFRGS), Porto Alegre, Brazil
\item \Idef{org71}Instituto de F\'{\i}sica, Universidad Nacional Aut\'{o}noma de M\'{e}xico, Mexico City, Mexico
\item \Idef{org72}iThemba LABS, National Research Foundation, Somerset West, South Africa
\item \Idef{org73}Johann-Wolfgang-Goethe Universit\"{a}t Frankfurt Institut f\"{u}r Informatik, Fachbereich Informatik und Mathematik, Frankfurt, Germany
\item \Idef{org74}Joint Institute for Nuclear Research (JINR), Dubna, Russia
\item \Idef{org75}Korea Institute of Science and Technology Information, Daejeon, Republic of Korea
\item \Idef{org76}KTO Karatay University, Konya, Turkey
\item \Idef{org77}Laboratoire de Physique Subatomique et de Cosmologie, Universit\'{e} Grenoble-Alpes, CNRS-IN2P3, Grenoble, France
\item \Idef{org78}Lawrence Berkeley National Laboratory, Berkeley, California, United States
\item \Idef{org79}Lund University Department of Physics, Division of Particle Physics, Lund, Sweden
\item \Idef{org80}Nagasaki Institute of Applied Science, Nagasaki, Japan
\item \Idef{org81}Nara Women{'}s University (NWU), Nara, Japan
\item \Idef{org82}National and Kapodistrian University of Athens, School of Science, Department of Physics , Athens, Greece
\item \Idef{org83}National Centre for Nuclear Research, Warsaw, Poland
\item \Idef{org84}National Institute of Science Education and Research, Homi Bhabha National Institute, Jatni, India
\item \Idef{org85}National Nuclear Research Center, Baku, Azerbaijan
\item \Idef{org86}National Research Centre Kurchatov Institute, Moscow, Russia
\item \Idef{org87}Niels Bohr Institute, University of Copenhagen, Copenhagen, Denmark
\item \Idef{org88}Nikhef, National institute for subatomic physics, Amsterdam, Netherlands
\item \Idef{org89}NRC Kurchatov Institute IHEP, Protvino, Russia
\item \Idef{org90}NRC «Kurchatov Institute»  - ITEP, Moscow, Russia
\item \Idef{org91}NRNU Moscow Engineering Physics Institute, Moscow, Russia
\item \Idef{org92}Nuclear Physics Group, STFC Daresbury Laboratory, Daresbury, United Kingdom
\item \Idef{org93}Nuclear Physics Institute of the Czech Academy of Sciences, \v{R}e\v{z} u Prahy, Czech Republic
\item \Idef{org94}Oak Ridge National Laboratory, Oak Ridge, Tennessee, United States
\item \Idef{org95}Ohio State University, Columbus, Ohio, United States
\item \Idef{org96}Petersburg Nuclear Physics Institute, Gatchina, Russia
\item \Idef{org97}Physics department, Faculty of science, University of Zagreb, Zagreb, Croatia
\item \Idef{org98}Physics Department, Panjab University, Chandigarh, India
\item \Idef{org99}Physics Department, University of Jammu, Jammu, India
\item \Idef{org100}Physics Department, University of Rajasthan, Jaipur, India
\item \Idef{org101}Physikalisches Institut, Eberhard-Karls-Universit\"{a}t T\"{u}bingen, T\"{u}bingen, Germany
\item \Idef{org102}Physikalisches Institut, Ruprecht-Karls-Universit\"{a}t Heidelberg, Heidelberg, Germany
\item \Idef{org103}Physik Department, Technische Universit\"{a}t M\"{u}nchen, Munich, Germany
\item \Idef{org104}Politecnico di Bari, Bari, Italy
\item \Idef{org105}Research Division and ExtreMe Matter Institute EMMI, GSI Helmholtzzentrum f\"ur Schwerionenforschung GmbH, Darmstadt, Germany
\item \Idef{org106}Rudjer Bo\v{s}kovi\'{c} Institute, Zagreb, Croatia
\item \Idef{org107}Russian Federal Nuclear Center (VNIIEF), Sarov, Russia
\item \Idef{org108}Saha Institute of Nuclear Physics, Homi Bhabha National Institute, Kolkata, India
\item \Idef{org109}School of Physics and Astronomy, University of Birmingham, Birmingham, United Kingdom
\item \Idef{org110}Secci\'{o}n F\'{\i}sica, Departamento de Ciencias, Pontificia Universidad Cat\'{o}lica del Per\'{u}, Lima, Peru
\item \Idef{org111}Shanghai Institute of Applied Physics, Shanghai, China
\item \Idef{org112}St. Petersburg State University, St. Petersburg, Russia
\item \Idef{org113}Stefan Meyer Institut f\"{u}r Subatomare Physik (SMI), Vienna, Austria
\item \Idef{org114}SUBATECH, IMT Atlantique, Universit\'{e} de Nantes, CNRS-IN2P3, Nantes, France
\item \Idef{org115}Suranaree University of Technology, Nakhon Ratchasima, Thailand
\item \Idef{org116}Technical University of Ko\v{s}ice, Ko\v{s}ice, Slovakia
\item \Idef{org117}Technische Universit\"{a}t M\"{u}nchen, Excellence Cluster 'Universe', Munich, Germany
\item \Idef{org118}The Henryk Niewodniczanski Institute of Nuclear Physics, Polish Academy of Sciences, Cracow, Poland
\item \Idef{org119}The University of Texas at Austin, Austin, Texas, United States
\item \Idef{org120}Universidad Aut\'{o}noma de Sinaloa, Culiac\'{a}n, Mexico
\item \Idef{org121}Universidade de S\~{a}o Paulo (USP), S\~{a}o Paulo, Brazil
\item \Idef{org122}Universidade Estadual de Campinas (UNICAMP), Campinas, Brazil
\item \Idef{org123}Universidade Federal do ABC, Santo Andre, Brazil
\item \Idef{org124}University of Cape Town, Cape Town, South Africa
\item \Idef{org125}University of Houston, Houston, Texas, United States
\item \Idef{org126}University of Jyv\"{a}skyl\"{a}, Jyv\"{a}skyl\"{a}, Finland
\item \Idef{org127}University of Liverpool, Liverpool, United Kingdom
\item \Idef{org128}University of Science and Techonology of China, Hefei, China
\item \Idef{org129}University of South-Eastern Norway, Tonsberg, Norway
\item \Idef{org130}University of Tennessee, Knoxville, Tennessee, United States
\item \Idef{org131}University of the Witwatersrand, Johannesburg, South Africa
\item \Idef{org132}University of Tokyo, Tokyo, Japan
\item \Idef{org133}University of Tsukuba, Tsukuba, Japan
\item \Idef{org134}Universit\'{e} Clermont Auvergne, CNRS/IN2P3, LPC, Clermont-Ferrand, France
\item \Idef{org135}Universit\'{e} de Lyon, Universit\'{e} Lyon 1, CNRS/IN2P3, IPN-Lyon, Villeurbanne, Lyon, France
\item \Idef{org136}Universit\'{e} de Strasbourg, CNRS, IPHC UMR 7178, F-67000 Strasbourg, France, Strasbourg, France
\item \Idef{org137}Universit\'{e} Paris-Saclay Centre d'Etudes de Saclay (CEA), IRFU, D\'{e}partment de Physique Nucl\'{e}aire (DPhN), Saclay, France
\item \Idef{org138}Universit\`{a} degli Studi di Foggia, Foggia, Italy
\item \Idef{org139}Universit\`{a} degli Studi di Pavia, Pavia, Italy
\item \Idef{org140}Universit\`{a} di Brescia, Brescia, Italy
\item \Idef{org141}Variable Energy Cyclotron Centre, Homi Bhabha National Institute, Kolkata, India
\item \Idef{org142}Warsaw University of Technology, Warsaw, Poland
\item \Idef{org143}Wayne State University, Detroit, Michigan, United States
\item \Idef{org144}Westf\"{a}lische Wilhelms-Universit\"{a}t M\"{u}nster, Institut f\"{u}r Kernphysik, M\"{u}nster, Germany
\item \Idef{org145}Wigner Research Centre for Physics, Hungarian Academy of Sciences, Budapest, Hungary
\item \Idef{org146}Yale University, New Haven, Connecticut, United States
\item \Idef{org147}Yonsei University, Seoul, Republic of Korea
\end{Authlist}
\endgroup
  %%%%%%% done by webmaster team
\end{document}